\documentclass[11pt,a4paper]{article}
\usepackage[utf8]{inputenc}
\usepackage[left=3cm,right=3cm,top=3cm,bottom=3cm]{geometry}

\usepackage{amsmath}
\usepackage{amsfonts}
\usepackage{amssymb}
\usepackage{mathtools}
\usepackage{cite}

\usepackage{color}
\definecolor{nicered}{rgb}{0.7,0.1,0.1}
\definecolor{nicegreen}{rgb}{0.1,0.5,.1}
\definecolor{mybrown}{RGB}{153,102,51}
 
\usepackage{graphicx}
\usepackage{subcaption}

\usepackage{multirow}

\usepackage{float}

\usepackage[colorlinks=true
,urlcolor=magenta
,anchorcolor=black
,citecolor=blue
,filecolor=black
,linkcolor=red
,menucolor=black
,linktocpage=true
,pdfproducer=medialab
]{hyperref}

\numberwithin{equation}{section}

\usepackage{authblk}

\title{Direct bounds on Left-Right gauge boson masses \\ at LHC Run~2} 

\date{}

\author[a]{Sergio~Ferrando~Solera}
\author[a]{Antonio~Pich}
\author[a]{Luiz~Vale~Silva}

\affil[a]{Departament de F\'{i}sica Te\`{o}rica, Instituto de F\'{i}sica Corpuscular,

Universitat de Val\`encia -- Consejo Superior de Investigaciones Cient\'{i}ficas,

Parc Cient\'{i}fic, Catedr\'{a}tico Jos\'{e} Beltr\'{a}n 2, E-46980 Paterna, Valencia, Spain}

\begin{document}

\hfill IFIC/23-28

{\let\newpage\relax\maketitle}

\maketitle


\textbf{Abstract.}
While the third run of the Large Hadron Collider (LHC) is ongoing, 
the underlying theory 
that extends the Standard Model remains so far unknown.
Left-Right Models (LRMs) introduce a new gauge sector, and can restore parity symmetry at high enough energies.
If LRMs are indeed realized in nature,
the mediators of the new weak force can be searched for in colliders via their direct production.
We recast existing experimental limits
from the LHC Run~2
and derive generic bounds on the masses of the heavy LRM gauge bosons. 
As a novelty, we discuss the 
dependence of the $W_R$ and $Z_R$ total width on the LRM scalar content,
obtaining model-independent bounds within the specific realizations of the LRM scalar sectors analysed here. These bounds avoid the need to detail the spectrum of the scalar sector, and apply in the general case where no discrete symmetry is enforced.
Moreover, we emphasize
the impact on the $W_R$ production at LHC of general textures of the right-handed quark mixing matrix without manifest left-right symmetry.
We find that the $W_R$ and $Z_R$ masses are constrained to lie above $2$~TeV and $4$~TeV, respectively.



\section{Introduction}

Having access to increasingly higher energies, physicists were able to observe the spectrum of fundamental particles portrayed by what became the Standard Model (SM).
After more than a decade of operation of the LHC, no Beyond the Standard Model (BSM) particles have been discovered so far.
Together with previous collider data, this can set important bounds on many BSM models, especially for models introducing new particles in the TeV range or below.
In many cases, such direct bounds compete or are better than indirect bounds derived from low-energy observables. An example would be processes involving the change of quark or lepton flavour that often probe very high-energy scales.
LHC will keep pushing the energy frontier to unprecedented levels during its new runs, and the possibility of unveiling new particles (scalars, spin-$ 1/\/2 $ fermions, vectors, and the like) presents an exciting prospect.
The mediators of new fundamental forces could then be discovered, and could hint at a deeper unification of particle physics interactions.
Left-Right Models (LRMs) could be one step towards such ultimate unification, and also prove to be an interesting candidate for BSM physics.

The starting point is the existence of new gauged interactions, extending the set of SM local symmetries to \cite{Mohapatra:1974hk,Mohapatra:1974gc}

\begin{equation}\label{eq:LRM_gauge}
    G_{LR} = SU(3)_{\rm QCD} \times SU(2)_L \times SU(2)_R \times U(1)_X
\end{equation}
where the quantum number associated with the Abelian symmetry $U(1)_X$ is baryon minus lepton numbers $B-L$.\footnote{Coincidentally, this is an Abelian symmetry that, along with the SM gauge group, leads to an anomaly-free theory in the presence of new SM-neutral fermions that carry a $B - L$ quantum number, such as right-handed neutrinos.}
The gauge couplings $g_L, g_R$ and $g_X$ of $SU(2)_L$, $ SU(2)_R $ and $U(1)_X$, respectively, will be assumed to be perturbatively small for all purposes here.
This larger symmetry group allows for the restoration of parity symmetry at some large energy scale \cite{Senjanovic:1975rk,Senjanovic:1978ev} (note that different definitions of parity symmetry are possible \cite{Ecker:1980at,Branco:1985ng}), which can provide new avenues for solving the strong $\mathcal{CP}$ problem \cite{Babu:1989rb,Babu:1988mw,Barr:1991qx,Lavoura:1996iu,Kuchimanchi:2010xs}. Alternatively, charge-conjugation symmetry could be restored at high energies, see Ref.~\cite{Maiezza:2010ic} for a phenomenological discussion.
The Left-Right (LR) gauge group in Eq.~\eqref{eq:LRM_gauge} can be embedded into a larger, more fundamental gauge group \cite{Pati:1974yy,Fritzsch:1974nn,Chang:1983fu,Chang:1984uy,Chang:1984qr}; the restoration of a discrete symmetry could then be pushed towards higher energy scales, implying that $ g_L \neq g_R $ at the characteristic energy scale $v_R$ associated with LRMs (see e.g. Ref.~\cite{Deppisch:2015cua}).
At the latter energy scale, lying much above the electroweak (EW) scale, the extra local symmetries are spontaneously broken, similar to the Brout-Englert-Higgs mechanism of the SM, resulting in the known gauge structure $ SU(3)_{\rm QCD} \times SU(2)_L \times U(1)_Y $; embedded in the larger group of symmetries of Eq.~\eqref{eq:LRM_gauge}, the hypercharge quantum number is $Y = T_{3R} + (B-L)/2$, where $ T_{3R} $ is the third generator of $SU(2)_R$, thus giving an origin for the particular values of hypercharge found in the SM.

The Spontaneous-Symmetry-Breaking (SSB) pattern $ SU(2)_R \times U(1)_X \to U(1)_Y $, followed by $ SU(2)_L \times U(1)_Y \to U(1)_{\rm EM} $, can be realized in multiple ways.
Here on, we discuss two scenarios of SSB, triggered (partially) by doublet $ (\mathbf{1}, \mathbf{1}, \mathbf{2})_{1/2} $ and $ (\mathbf{1}, \mathbf{2}, \mathbf{1})_{1/2} $ \cite{Pati:1974yy,Mohapatra:1974hk,Mohapatra:1974gc,Senjanovic:1975rk,Mohapatra:1976iz,Mohapatra:1977mj}, or by triplet $ (\mathbf{1}, \mathbf{1}, \mathbf{3})_1 $ and $ (\mathbf{1}, \mathbf{3}, \mathbf{1})_1 $ \cite{Mohapatra:1979ia,Mohapatra:1980yp}, scalar fields, where the charge in subscript is $(B-L)/2$.


As in the case of Left-Handed (LH) fermions,
Right-Handed (RH) quark and lepton fields fill doublet irreducible representations, which requires the introduction of RH neutrinos in the latter case.
This means that weak interactions treat quarks and leptons on an equal footing, and neutrinos can acquire a mass from renormalizable interactions.
For instance, different fermions can acquire masses via Yukawa interactions with a bi-doublet scalar field $ (\mathbf{1}, \mathbf{2}, \mathbf{2})_0 $.
Moreover, in the triplet scenario, Majorana mass terms for the neutrinos become possible \cite{Mohapatra:1979ia,Mohapatra:1980yp} (see e.g. Refs.~\cite{Mohapatra:1986uf,Mohapatra:1998rq} for a discussion about light neutrino mass stability). 
The bi-doublet scalar also participates in the pattern of SSB
and, to avoid large BSM contributions to the EW $ \rho $ parameter already at the tree level,\footnote{At the tree level in the SM $ \rho \coloneqq M_W^2 / (M_Z^2 \, \cos^2 \theta_W) = 1 $, $ \theta_W $ being the weak angle with $ \cos^2 \theta_W = g_L^2 / (g_L^2 + g_Y^2) $.} the vacuum expectation value (VEV) of the triplet scalar charged under $ SU(2)_L $ is set to a value much below the EW scale (for the stability of this condition, see e.g. Refs.~\cite{Mohapatra:1986uf,Mohapatra:1998rq}), in which case the bi-doublet becomes the solely responsible for the EW SSB in the triplet scenario.
In the doublet scenario, however, it is possible that the EW transition is triggered mostly by the doublet scalar charged under $ SU(2)_L $ \cite{Bernard:2020cyi,Karmakar:2022iip}.
The resulting SM-like Higgs particle $h^0$ is an admixture of the scalar field content discussed above.

For various theoretical reasons, larger scalar sectors have also been considered, e.g., to separate the restoration of the LR
gauge group from that of parity symmetry \cite{Chang:1983fu,Chang:1984uy}.
An interesting alternative for the mechanism of fermion mass generation is the introduction of
vector-like fermions (whose mass terms do not depend on the scalar sector, nor their scale must be related to the EW scale) 
in such a way that the masses of the SM-like fermions are generated via Yukawa terms with doublet scalar fields \cite{Davidson:1987mh,Rajpoot:1986nv,Rajpoot:1987fca,Chang:1986bp,Mohapatra:1987nx,Babu:1988yq,Babu:1988mw,Babu:1989rb,Balakrishna:1987qd,Balakrishna:1988ks,Balakrishna:1988bn}. Several recent papers have discussed phenomenological implications of this possibility, e.g., Refs.~\cite{Babu:2018vrl,Babu:2022ikf,Dey:2022tbp,Dcruz:2022rjg,Dcruz:2023mvf,Babu:2023srr}.

As in the SM, the diagonalization of the quark mass matrix introduces an SM-like Cabibbo-Kobayashi-Maskawa (CKM) matrix 
in the charged-current interactions of LH quarks, together with a counterpart matrix 
in the charged-current interactions of RH quarks; a similar comment also holds for the leptonic sector.
These mixing matrices can be related if a discrete symmetry is enforced, producing the well-known manifest and pseudo-manifest LR symmetry cases, see e.g. Refs.~\cite{Zhang:2007fn,Senjanovic:2014pva,Senjanovic:2015yea}.
However, we will not consider that such a symmetry is realized at the 
LR SSB  energy scale, and will not assume a particular structure for the RH mixing matrices, thus including in our investigation textures with non-manifest LR symmetry. Such matrices can be constrained by flavour observables \cite{Langacker:1989xa,Zhang:2007da,Blanke:2011ry,Dekens:2021bro}, such as $ K^0$-$ \overline{K}^0 $ neutral meson mixing \cite{Bertolini:2014sua,Bernard:2015boz}, that exceed the scope of the present work.

A lot of attention has been dedicated to the search of new heavy vector $Z', W'$ bosons of different origins. 
Past collider experiments such as LEP-II and the Tevatron searched for the direct production of these new states, excluding 
heavy gauge bosons with masses slightly above the EW scale
(for SM-like new gauge-coupling values) \cite{Workman:2022ynf}.
These past collider bounds have been significantly improved at the LHC; among the available experimental analyses,
we highlight the following searches:
$Z'$ decaying to a lepton pair \cite{ATLAS:2017eiz,ATLAS:2019erb,CMS:2021ctt}; di-jet resonances \cite{ATLAS:2019fgd,CMS:2019gwf}; di-boson decay modes \cite{ATLAS:2019nat,ATLAS:2020fry,CMS:2021fyk,CMS:2021klu}; and $W'$ leptonic decays \cite{ATLAS:2019lsy,CMS:2022yjm,ATLAS:2021bjk,CMS:2018fza}.
Some collider bounds apply directly to the so-called Sequential SM
\cite{Altarelli:1989ff}, that is a benchmark for the sensitivity of direct searches in discovering new gauge bosons.
On the other hand,
there are bounds that apply directly to LRMs in the presence of heavy RH neutrinos \cite{CMS:2018iye,ATLAS:2019isd,CMS:2021dzb,ATLAS:2023cjo},
or a $ t \bar{b} $ quark pair in the final state (with subsequently $ t \to W^+ b $, with $ W^+ \to \nu_\ell \ell^+$ 
or $ W^+ \to q \bar{q}' $, where $ \nu_\ell $ is a light mostly-LH neutrino) \cite{ATLAS:2018uca,ATLAS:2018wmg,ATLAS:2023ibb,CMS:2021mux,CMS:2023gte}, see also Refs.~\cite{Sullivan:2002jt,Duffty:2012rf}.
Note that the latter bound strongly depends on the specific coupling of the charged LRM gauge boson to $ t \bar{b} $, while the values of RH neutrino masses are in principle unknown.
It is worth stressing the intensive research program for scrutinizing the parameter space of LRMs with
massive neutrinos in colliders, that resulted in fruitful proposals, see e.g. Refs.~\cite{Keung:1983uu,Chakrabortty:2012pp,Nemevsek:2018bbt,Frank:2023epx}.
For a recent reference on prospects for a future collider candidate, see Ref.~\cite{Nemevsek:2023hwx}.

In this article,
we explore direct bounds that can be set on $Z' \coloneqq Z_R$ and $W' \coloneqq W_R$ masses, which are the heavy LRM counterparts of the SM-like $Z \coloneqq Z_L$ and $W \coloneqq W_L$ gauge bosons.\footnote{Note that from this point on the SM-like gauge bosons will be called $Z_L$ and $W_L$, while the new gauge bosons will be denoted $Z_R$ and $W_R$.}
Bounds on the Sequential SM
cannot be directly interpreted in terms of LRMs, but also existing direct searches of heavy LRM gauge bosons do not entirely reflect the freedom in the parameter space of LRMs. It is often assumed that the  
RH gauge coupling or the RH mixing matrix in the quark sector are equal to their SM counterparts, or that the new gauge bosons do not couple to scalar or vector bosons in searches involving fermions in the final state.
Given the variety of ways in which the scalar sector of LRMs can be realized, we focus on final states consisting of fermion pairs,
for which the particular scalar sector manifests only in the total width of the new gauge bosons, while the couplings of the latter to fermions maintain the 
same structure (the fermion mixing matrices may change however according to the case analysed).
We then exploit experimental results valid for different total widths of the decaying heavy particles. In the cases discussed hereafter, we find widths smaller than
10\% of the heavy gauge-boson masses.
Interference effects of the $W_L$ with the $W_R$ are mixing suppressed, or chiral suppressed at production and/or decay since the $W_R$ couples predominantly to RH fermions; for a discussion of interference terms in the context of the SSM, see Refs.~\cite{Rizzo:2007xs,Accomando:2011eu}, and references therein.
We neglect interference terms of the $Z_R$ with the SM neutral gauge bosons, which is valid as long as we remain close enough to the resonance peak; for discussions, see Refs.~\cite{Accomando:2010fz,Accomando:2013sfa} and also Refs.~\cite{ATLAS:2017fih,CMS:2021ctt}.

We illustrate the impact of the scalar sector by providing bounds for the following specific realizations:
(D) a model with two doublets and a bi-doublet; (T) a model with two triplets and a bi-doublet; and (Eff) an effective model with two doublets and no bi-doublet, in which fermion masses are generated via non-renormalizable dimension-five operators. Note that we need the bi-doublet in scenario (T) to trigger EW SSB.
For the sake of avoiding any further detailing of the specific realization of the scalar sector,
when calculating the total widths of the heavy gauge bosons,
we consider the limit in which the extended scalar sector is light, i.e., kinematically accessible. 
This allows then to establish model-independent bounds in scenarios (D), (T) and (Eff).
This procedure, which translates into the largest total widths in these three scenarios for a fixed $g_R$ coupling, also leads to the less constraining lower bounds on the heavy gauge-boson masses, mainly due to branching ratios into fermion pairs becoming smaller.
This strategy does not alleviate much the lower bounds obtained with respect to the limits quoted by the experimental collaborations.
We also discuss how the bounds depend on the coupling $g_R$ of the new gauge sector and the structure of the RH counterpart of the CKM matrix.
We focus on the extreme case of light RH neutrinos that gives the more conservative bounds,
since the $ Z_R $ and $ W_R $ total widths are larger in this case,
and reserve the discussion of the mixing matrix in the neutrino sector to future work, as in the case of RH quarks.



This article is organized as follows: in Sec.~\ref{sec:models} we provide details about the LRMs considered; in Sec.~\ref{sec:direct_bounds} we establish bounds on distinct versions of LRMs, $ Z_R $ and $ W_R $ being the main focus of subsections \ref{sec:direct_bounds_Zprime} and \ref{sec:direct_bounds_Wprime}, respectively, while Sec.~\ref{sec:mass_relation} discusses
a summary of direct bounds, and
the 
improved limits
that can be derived from the use of the relation between these two masses, and the bounds on the LR energy scale $v_R$ 
obtained in the different realizations of the scalar sector.
Conclusions are found in Sec.~\ref{sec:conclusions}.
In App.~\ref{sec:Scalar_Potentials} we give the scalar potentials in scenarios (D) and (T); the relevant couplings are collected in App.~\ref{sec:summary_couplings}; finally, App.~\ref{sec:Non-Fermionic_Decays} contains a more detailed discussion about $Z_R$ and $W_R$ decays into two bosons.


\section{The models}\label{sec:models}




We first discuss some generic aspects of LRMs.
In the weak sector, we have two charged and three neutral gauge bosons, besides 
three different coupling constants.
Typically, in LRMs we find a mixing between the \(W_L\) and \(W_R\), and between the \(Z_L\) and \(Z_R\) gauge bosons. Defining \(W_{1,2}^{-\mu}\coloneqq\left(W_{L,R}^{1\mu}+iW_{L,R}^{2\mu}\right)/\sqrt{2}\), the mass eigenstates \(W_L^{-\mu}\) and \(W_R^{-\mu}\) are obtained using the unitary transformation

\begin{equation}
    \label{ecu.2.1}
    \left(\!\!
    \begin{array}{c}
        W_L^{-\mu} \\
        W_R^{-\mu}
    \end{array}\!\!\right)=\left(
    \begin{array}{cc}
    \cos\xi & -e^{-i\lambda}\sin\xi  \\
    e^{i\lambda}\sin\xi  & \cos\xi
    \end{array}\right)\left(\!\!
    \begin{array}{c}
    W_1^{-\mu} \\
    W_2^{-\mu}
    \end{array}\!\!\right),
\end{equation}

\noindent where \(\lambda\) and \(\xi\) are fixed by the VEVs of the scalar fields. On the other hand, for the neutral gauge bosons we have

\begin{equation}
    \label{ecu.2.2}
    \left(\!\!
    \begin{array}{c}
    Z_L^{\mu} \\
    Z_R^{\mu} \\
    A^{\mu}
    \end{array}\!\!\right)=\left(
    \begin{array}{ccc}
    c_\alpha & -s_\alpha & 0 \\
    s_\alpha & c_\alpha & 0 \\
    0 & 0 & 1
    \end{array}\right)\left(
    \begin{array}{ccc}
    c_{\theta_W} & 0 & -s_{\theta_W} \\
    0 & 1 & 0 \\
    s_{\theta_W} & 0 & c_{\theta_W}
    \end{array}\right)\left(
    \begin{array}{ccc}
    1 & 0 & 0 \\
    0 & c_\gamma & -s_\gamma  \\
    0 & s_\gamma & c_\gamma
    \end{array}\right)\left(\!\!
    \begin{array}{c}
    W_L^{3\mu} \\
    W_R^{3\mu} \\
    W_{X}^{\mu}
    \end{array}\!\!\right),
\end{equation}

\noindent
where $ c_\alpha \equiv \cos \alpha $, etc.
The mixing parameters \(\sin\xi\) and \(\sin\alpha\) will depend on the particular SSB mechanism in use. Nonetheless, they are usually proportional to \(v_{EW}^2/v_{R}^2\), and thus small.


In the breaking of \(SU(2)_R\times U(1)_{X}\) into \(U(1)_Y\) we needed to introduce in Eq.~\eqref{ecu.2.2} a mixing angle \(\gamma\) between the \(W_{X}\) and \(W_R^3\) fields completely analogous to the weak angle \(\theta_W\) in the SM. Then, if we want to reproduce the phenomenology of the EW theory at low energies we need to impose the condition

\begin{equation}
    \label{ecu.2.3}
    e=g_L\sin\theta_W=g_R\sin\gamma\cos\theta_W=g_{X}\cos\gamma\cos\theta_W.
\end{equation}

\noindent
From the previous equation we can set bounds on the allowed values of the angle \(\gamma\). Assume that we want to stay in the perturbative regime, namely, we take \(g_R,\,g_{X}<1\). Then, we arrive at the condition

\begin{equation}
    \label{ecu.2.4}
    \sqrt{1-\frac{e^2}{\cos^2\theta_W}}>\sin\gamma>\frac{e}{\cos\theta_W}.
\end{equation}

\noindent
Therefore, \(70^{\circ}\gtrsim\gamma\gtrsim20^{\circ}\).
The parity symmetric scenario $g_L = g_R$ corresponds to \(\sin\gamma=\tan\theta_W\), so \(\gamma\approx33^{\circ}\).
Nevertheless, as we have pointed out in the introduction, LRMs can be embedded in theories with a larger gauge group,
and in general we will not assume neither \(\mathcal{P}\) nor \(\mathcal{C}\) symmetry in this work. The running of the coupling constants produces a difference between \(g_L\) and \(g_R\) at the LR scale $v_R$, giving different values for $\gamma$ satisfying the perturbativity constraints above. We will not discuss embeddings leading to a particular value of $\gamma$.


The matter content of the LRMs considered here consists of the doublets \(q_{L,\,R}^m\coloneqq\left(u_{L,\,R}^m,\,d_{L,\,R}^m\right)^{\mathrm{T}}\) and \(\ell_{L,\,R}^m\coloneqq\left(\nu_{L,\,R}^m,\,e_{L,\,R}^m\right)^{\mathrm{T}}\), where we use the index \(m\) for the three different families of fermions. They transform as \(q_L^m\sim\left(\mathbf{3},\mathbf{2},\mathbf{1}\right)_{1/6}\), \(q_R^m\sim\left(\mathbf{3},\mathbf{1},\mathbf{2}\right)_{1/6}\), \(\ell_L^m\sim\left(\mathbf{1},\mathbf{2},\mathbf{1}\right)_{-1/2}\) and \(\ell_R^m\sim\left(\mathbf{1},\mathbf{1},\mathbf{2}\right)_{-1/2}\) under the gauge group \(G_{LR}\). The covariant derivatives of the fermions are

\begin{equation}
    \label{ecu.2.5}
    D^{\mu}\,f_{L,R}^{m}\coloneqq \left\{\partial^{\mu}+i\,g_{L,R}\,\dfrac{\sigma^{k}}{2}\,W_{L,R}^{k\mu}+i\,g_{X}\,\frac{B-L}{2}\,W_{X}^{\mu}\right\}f_{L,R}^{m},
\end{equation}

\noindent where  \(\sigma^k\) are the Pauli matrices and \(f=q,\,\ell\).


\subsection{The bi-doublet field}


To start the discussion of the SSB of the LR gauge group, we introduce a bi-doublet field \(\Phi\sim\left(\mathbf{1},\mathbf{2},\mathbf{2}\right)_{0}\), together with its charge-conjugated \(\tilde{\Phi}\coloneqq\sigma^2\Phi^{\ast}\sigma^2\sim\left(\mathbf{1},\mathbf{2},\mathbf{2}\right)_{0}\).
With the charge operator \(\mathcal{Q}\coloneqq\left(\sigma^3/2\right)_L+\left(\sigma^3/2\right)_R+\left(B-L\right)/2\), the bi-doublet can be written in terms of its components as

\begin{equation}
    \label{ecu.2.6}
    \Phi=\left(
    \begin{array}{cc}
    \phi_1^0 & \phi_1^+ \\
    \phi_2^- &  \phi_2^0
    \end{array}\right).
\end{equation}

\noindent Its VEV takes the form 

\begin{equation}
    \label{ecu.2.7}
    \left<\Phi\right>=\frac{1}{\sqrt{2}}\left(
    \begin{array}{cc}
    \kappa_1 & 0\\
    0 & \kappa_2
    \end{array}\right)
\end{equation}


\noindent
where $ \kappa_{1,2} $ carry in principle complex phases \cite{Zhang:2007da}.
Being neutral under $U(1)_{X}$, we introduce other fields to break the LR gauge group into the low-energy gauge group $ SU(3)_{\rm QCD} \times U(1)_{\rm EM} $, and to set a large gap between the LR and EW scales.

Besides producing masses to the gauge bosons, fermion mass terms can also be generated.
The Yukawa term for quarks is simply

\begin{equation}
    \label{ecu.2.8}
    \mathcal{L}_Y^q=-\sum_{m,\,n}\bar{q}_{L}^{\,m}\left(r_{mn}\Phi+s_{mn}\tilde{\Phi}\right)q_{R}^{\,n}+\mathrm{h.c.},
\end{equation}

\noindent where \(r\) and \(s\) are two completely general complex matrices.
Under \(\mathcal{P}:\Phi\rightarrow\Phi^{\dagger}\), and thus imposing parity symmetry would imply that \(r\) and \(s\) are hermitian matrices, while under \(\mathcal{C}:\Phi\rightarrow\Phi^{\ast}\), and charge-conjugation invariance would imply that they are real.

In order to diagonalize the mass matrices of quarks we need to introduce the unitary matrices \(V_L^{\mathrm{CKM}}\) and \(V_{R}^{\mathrm{CKM}}\). We have the following Lagrangian for the charged currents mediated by the \(W\) bosons:

\begin{equation}
    \label{ecu.2.9}
    \mathcal{L}_{CC}^q=-\frac{1}{\sqrt{2}}\bar{u}\,\gamma_{\mu}\left(g_LW_{1}^{+\mu}V_{L}^{\textrm{CKM}}\mathcal{P}_{L}+g_RW^{+\mu}_{2}V_{R}^{\textrm{CKM}}\mathcal{P}_{R}\right)d+\mathrm{h.c.},
\end{equation}

\noindent where \(u\coloneqq\left(u,c,t\right)^{\rm T}\) and \(d\coloneqq\left(d,s,b\right)^{\rm T}\). We can make a redefinition of the fields in order to write \(V_{L}^{\mathrm{CKM}}\) in the same way as the CKM matrix of the SM. On the other hand, \(V_R^{\mathrm{CKM}}\) will remain a completely general unitary matrix. (Nonetheless, if we impose \(\mathcal{CP}\) symmetry, \(V_L^{\mathrm{CKM}}=V_R^{\mathrm{CKM}}\).)


The Yukawa term of Eq.~(\ref{ecu.2.8}) generates FCNCs mediated by tree-level scalar exchanges.
We note that this does not occur in the gauge sector when the mechanism for generating fermion masses is as above, since fermions come in sequential generations of identical quantum numbers \cite{Glashow:1970gm} (such would not be the case when having vector-like fermions as in Refs.~\cite{Babu:2018vrl,Dcruz:2022rjg,Dcruz:2023mvf}). Explicitly, the interaction Lagrangian among quarks and the \(Z_R\) boson is

\begin{equation}
    \label{ecu.2.10}
    \begin{split}
    \mathcal{L}^q_{RNC}&=\frac{1}{12}\frac{e}{\cos\theta_W}\tan\gamma\,Z_R^{\mu}\,\bar{u}\gamma_{\mu}\left\{\left(2-3\cot^2\gamma\right)-\left(3\cot^2\gamma\right)\gamma_5\right\}u
        \\& +\frac{1}{12}\frac{e}{\cos\theta_W}\tan\gamma\,Z_R^{\mu}\,\bar{d}\gamma_{\mu}\left\{\left(2+3\cot^2\gamma\right)+\left(3\cot^2\gamma\right)\gamma_5\right\}d\, ,
    \end{split}
\end{equation}

\noindent
neglecting the \(Z_L-Z_R\) mixing;
the couplings among quarks and the \(Z_L\) gauge boson are then the ones of the SM.


\subsection{A model with one bi-doublet, plus two doublet fields (D)}


In order to break spontaneously \(SU(2)_R\times U(1)_{X}\) into \(U(1)_Y\) at a high-energy scale much above the EW scale we still need to add more scalar degrees of freedom. The most common choices are either two doublets or two triplets (although, since we are not focused in restoring parity symmetry at the LRM scale $v_R$, we could consider the case where we simply add one doublet or one triplet). Using doublets, we introduce \(\chi_L\sim\left(\mathbf{1},\mathbf{1},\mathbf{2}\right)_{1/2}\) and \(\chi_R\sim\left(\mathbf{1},\mathbf{2},\mathbf{1}\right)_{1/2}\). They are written in terms of their components as

\begin{equation}
\label{ecu.2.11}
\chi_{L,R}=\left(
\begin{array}{c}
\chi_{L,R}^+ \\
\chi_{L,R}^0
\end{array}\right) ,
\end{equation}

\noindent whose VEVs are
\begin{equation}
    \label{ecu.2.12}
    \left<\chi_{L,R}\right>=\frac{1}{\sqrt{2}}\left(
    \begin{array}{c}
    0 \\
    v_{L,R}
    \end{array}\right)
\end{equation}

\noindent where $ v_{L,R} $ carry in principle complex phases \cite{Zhang:2007da}. For phenomenological reasons we need to impose the condition \(\left|v_R\right|\gg\left|\kappa_1\right|,\left|\kappa_2\right|\) and \(\left|v_L\right|\). At leading order in \(v_{EW}^2/v_R^2\) the masses of the gauge bosons are

\begin{equation}
    \label{ecu.2.13}
    \begin{array}{ll}
    M_{W_L}^2=\dfrac{1}{4}g_L^2\left\{\left|\kappa_1\right|^2+\left|\kappa_2\right|^2+\left|v_L\right|^2\right\}, &
    M^2_{W_R}=\dfrac{1}{4}g_R^2\left|v_R\right|^2, \\
    M_{Z_L}^2=\dfrac{g_L^2g_R^2+g_L^2g_{X}^2+g_R^2g_{X}^2}{4\left(g_{X}^2+g_R^2\right)}\left\{\left|\kappa_1\right|^2+\left|\kappa_2\right|^2+\left|v_L\right|^2\right\}, &
    M_{Z_R}^2=\dfrac{1}{4}\left(g_R^2+g_{X}^2\right)\left|v_R\right|^2. \\
    \end{array}
\end{equation}

\noindent
As stated earlier, in these models \(v_{EW}^2=\left|\kappa_1\right|^2+\left|\kappa_2\right|^2+\left|v_L\right|^2\) and, at leading order in \(v_{EW}^2/v_R^2\), \(\rho = M_{W_L}^2/(M_{Z_L}^2\cos^2\theta_W)=1\).


In the present case we have six neutral and two singly-charged physical scalar fields. Only one of the neutral fields has a mass proportional to the EW scale, which corresponds to the Higgs boson of the SM.


We cannot use doublets to produce Yukawa terms of dimension four. Thus, the mass term for leptons is completely analogous to the one of quarks in Eq.~(\ref{ecu.2.8}). Consequently, we also need to introduce the $ 3 \times 3 $ unitary matrices \(V_L^{\mathrm{PMNS}}\) and \(V_R^{\mathrm{PMNS}}\). Again, we can make a redefinition of the fields to write \(V_L^{\mathrm{PMNS}}\) as in the SM (i.e., the mixing matrix measured in neutrino oscillation experiments), while \(V_R^{\mathrm{PMNS}}\) is a completely general unitary matrix. Charged-current interactions of leptons mediated by the \(W\) bosons are described by

\begin{equation}
    \label{ecu.2.14}
    \mathcal{L}_{CC}^\ell=-\frac{1}{\sqrt{2}}\bar{\nu}\,\gamma_{\mu}\left(g_LW_{1}^{+\mu}V_{L}^{\textrm{PMNS}}\mathcal{P}_{L}+g_RW^{+\mu}_{2}V_{R}^{\textrm{PMNS}}\mathcal{P}_{R}\right)e+\mathrm{h.c.}
\end{equation}

\noindent On the other hand, the interaction Lagrangian for the \(Z_R\) boson and leptons is 

\begin{equation}
    \label{ecu.2.15}    
    \begin{split}
    \mathcal{L}^\ell_{RNC}&=-\frac{1}{4}\frac{e}{\cos\theta_W}\tan\gamma\,Z_R^{\mu}\,\bar{e}\gamma_{\mu}\left\{\left(2-\cot^2\gamma\right)-\left(\cot^2\gamma\right)\gamma_5\right\}e
    \\& -\frac{1}{2}\frac{e}{\cos\theta_W}\tan\gamma\, Z_R^{\mu}\,\bar{\nu}_L\gamma_{\mu}\nu_L-\frac{1}{2}\frac{e}{\cos\theta_W}\tan\gamma\left(1+\cot^2\gamma\right)\,Z_R^{\mu}\,\bar{\nu}_R\gamma_{\mu}\nu_R.
    \end{split}
\end{equation}


\subsection{A model with one bi-doublet, plus two triplet fields (T)}


The model with doublets does not explain why LH neutrinos are so much more lighter than the other fermions. In order to give an answer to this problem a model with triplets can be proposed instead. Here we introduce \(\Delta_L\sim\left(\mathbf{1},\mathbf{3},\mathbf{1}\right)_{1}\) and \(\Delta_R\sim\left(\mathbf{1},\mathbf{1},\mathbf{3}\right)_{1}\), whose expressions in terms of their components can be put under the form

\begin{equation}
    \label{ecu.2.16}
    \Delta_{L,R}=\left(
    \begin{array}{cc}
    \delta_{L,R}^+/\sqrt{2}     &  \delta^{++}_{L,R} \\
    \delta_{L,R}^0     &  -\delta_{L,R}^+/\sqrt{2}
    \end{array}
    \right).
\end{equation}

\noindent Their VEVs are

\begin{equation}
    \label{ecu.2.17}
    \left<\Delta_{L,R}\right>=\frac{1}{\sqrt{2}}\left(
    \begin{array}{cc}
    0 & 0 \\
    v_{L,R} & 0 
    \end{array}\right) ,
\end{equation}

\noindent where $ v_{L,R} $ carry in principle complex phases \cite{Zhang:2007da}. The masses of the gauge bosons are 

\begin{equation}
    \label{ecu.2.18}
    \begin{array}{ll}
    M_{W_L}^2=\dfrac{1}{4}g_L^2\left\{\left|\kappa_1\right|^2+\left|\kappa_2\right|^2+2\left|v_L\right|^2\right\}, &
    M^2_{W_R}=\dfrac{1}{2}g_R^2\left|v_R\right|^2, \\
    M_{Z_L}^2=\dfrac{g_L^2g_R^2+g_L^2g_{X}^2+g_R^2g_{X}^2}{4\left(g_{X}^2+g_R^2\right)}\left\{\left|\kappa_1\right|^2+\left|\kappa_2\right|^2+4\left|v_L\right|^2\right\}, &
    M_{Z_R}^2=\left(g_R^2+g_{X}^2\right)\left|v_R\right|^2. \\
    \end{array}
\end{equation}

\noindent
If we want to preserve the condition \(\rho\approx1\), it is necessary to assume that \(\left|v_R\right|^2\gg\left|\kappa_1\right|^2+\left|\kappa_2\right|^2\gg\left|v_L\right|^2\).


The models with triplets contain six neutral, two singly-charged and two doubly-charged scalars.
Only one of the neutral fields has a mass proportional to the EW scale, again corresponding to the Higgs boson of the SM.


As stated previously, when having triplet fields we can introduce the following Majorana mass term for leptons

\begin{equation}
    \label{ecu.2.19}
    \mathcal{L}_M=-\sum_{m,n}\left\{\bar{\ell}_R^{\,c\,m}\left(h_R\right)_{mn}\,i\sigma^2\Delta_R\,\ell_R^{\,n}+\bar{\ell}_L^{\,c\,m}\left(h_L\right)_{mn}\,i\sigma^2\Delta_L\,\ell_L^{\,n}\right\}+\textrm{h.c.}
\end{equation}

\noindent
Therefore, the masses of the LH neutrinos can be explained via a seesaw mechanism, namely, the masses of the heavy neutrinos \(\nu_h\) are proportional to the LR scale \(v_R\), while the masses of the light ones \(\nu_l\) are proportional to \(v_{EW}^2/v_R\). In the limit \(M_{\nu_h}\gg M_{\nu_l}\), we can diagonalize their mass matrices introducing the $3 \times 3$ unitary matrices \(V_{l}^{\mathrm{PMNS}}\) and \(V_{h}^{\mathrm{PMNS}}\). Since lepton number is not a conserved quantity anymore due to the interaction term Eq.~(\ref{ecu.2.19}), the matrix \(V_l^{\mathrm{PMNS}}\) will contain two additional free parameters (the Majorana phases) when compared to \(V_L^{\mathrm{CKM}}\). In this limit \(\nu_l\approx\nu_L\), \(\nu_h\approx\nu_R\) and the charged and neutral currents are given by the Lagrangians in Eqs.~(\ref{ecu.2.14}) and (\ref{ecu.2.15}), changing \(V_{L,R}^{\mathrm{PMNS}}\) and \(\nu_{L,R}\) by \(V_{l,h}^{\mathrm{PMNS}}\) and \(\nu_{l,h}\), respectively.


\subsection{An effective model with two doublet fields (Eff)}


The scalar potentials for both doublets and triplets are shown in App.~\ref{sec:Scalar_Potentials}. As seen therein, these theories contain a huge amount of free parameters.
We now move to another LRM with an extremely simple scalar sector, which we refer to as the \(\chi_L+\chi_R\) Effective LR Model.
In this case the leading picture of SSB consists of \(SU(2)_R\times U(1)_X\) being broken into \(U(1)_Y\) by the doublet \(\chi_R\) and \(SU(2)_L\times U(1)_Y\) being broken into \(U(1)_{\rm EM}\) by \(\chi_L\). We can always work in the unitary gauge, where 

\begin{equation}
    \label{ecu.2.20}
    \chi_{L,R}=\frac{1}{\sqrt{2}}\left(
    \begin{array}{c}
    0 \\
    v_{L,R}+\chi_{L,R}^{0r}
    \end{array}
    \right)
\end{equation}

\noindent with \(v_{L,R}\) being two real parameters and \(\chi_{L,R}^{0r}\) two hermitian fields. The scalar self-interaction is described by

\begin{equation}
    \label{ecu.2.21}
    V=-\mu_L^2\chi_L^{\dagger}\chi_L-\mu_R^2\chi_R^{\dagger}\chi_R+\lambda_L\left(\chi_L^{\dagger}\chi_L\right)^2+\lambda_R\left(\chi_R^{\dagger}\chi_R\right)^2+\lambda_{LR}\left(\chi_L^{\dagger}\chi_L\right)\left(\chi_R^{\dagger}\chi_R\right).
\end{equation}

All of the parameters above must be real in order to have a hermitian potential. Imposing that it must be bounded from below, we get that \(\lambda_L>0\), \(\lambda_R>0\) and \(4\lambda_L\lambda_R-\lambda_{LR}^2>0\). The VEVs are obtained from solving

\begin{equation}
    \label{ecu.2.22}
    \left\{
    \begin{array}{c}
    v_L\left(-\mu_L^2+\lambda_Lv_L^2+\lambda_{LR}v_R^2/2\right)=0 \\
    \\
    v_R\left(-\mu_R^2+\lambda_Rv_R^2+\lambda_{LR}v_L^2/2\right)=0
    \end{array}\right. .
\end{equation}

In contrast to the other two cases, if we want the condition \(v_R\gg v_L\neq0\) to be satisfied, we cannot impose parity symmetry in the potential.\footnote{Note that the (explicit) breaking of parity can be achieved with soft terms.}


In this  
model there is no mixing between the \(W_L\) and \(W_R\) bosons at the tree level, a mixing being generated at one loop. They are then mass eigenstates at the order of trees. The masses of the gauge fields are

\begin{equation}
    \label{ecu.2.23}
    \begin{array}{ll}
    M_{W_L}=\dfrac{1}{2}g_Lv_L, & M_{Z_L}^2=\dfrac{g_L^2g_R^2+g_L^2g_{X}^2+g_R^2g_{X}^2}{4\left(g_X^2+g_R^2\right)}v_L^2, \\
    M_{W_R}=\dfrac{1}{2}g_Rv_R, & M_{Z_R}^2=\dfrac{1}{4}\left(g_R^2+g_{X}^2\right)v_R^2. 
    \end{array}
\end{equation}

\noindent
The EW scale is just \(v_L\) at this order. Conversely, there is a mixing between the \(Z_L\) and \(Z_R\) gauge bosons given by the angle

\begin{equation}
    \label{ecu.2.24}
    \tan\left(2\alpha\right)=\frac{2g_{X}^2\sqrt{g_L^2g_{X}^2+g_{R}^2g_{X}^2+g_R^2g_L^2}\,v_L^2}{\left(g_R^2+g_{X}^2\right)^2v_R^2-\left(g_L^2g_{X}^2+g_{R}^2g_{X}^2+g_R^2g_L^2-g_{X}^4\right)v_L^2}.
\end{equation}


\noindent The physical Higgs fields \(h\) and \(H\) are obtained using the transformation

\begin{equation}
    \label{ecu.2.25}
    \left(
    \begin{array}{c}
    H \\
    h
    \end{array}\right)=\left(\begin{array}{cc}
    \cos\theta & -\sin\theta \\
    \sin\theta & \cos\theta
    \end{array}\right)\left(\begin{array}{c}
    \chi_R^{0r} \\
    \chi_L^{0r}
    \end{array}\right)
\end{equation}

\noindent with 

\begin{equation}
    \label{ecu.2.26}
    \tan\left(2\theta\right)=\frac{v_Lv_R\lambda_{LR}}{\lambda_Lv_L^2-\lambda_Rv_R^2}.
\end{equation}

\noindent At leading order in \(v_L^2/v_R^2\), their masses are given by

\begin{equation}
    \label{ecu.2.27}
    M^2_{H}\approx 2\lambda_R v_R^2,\hspace{0.8cm}M^2_{h}\approx\frac{4\lambda_L\lambda_R-\lambda_{LR}^2}{2\lambda_R}\,v_L^2.
\end{equation}


It is clear that we cannot use dimension-four operators to generate mass terms for fermions using only doublets. Instead, in this model their masses can be generated by effective operators of dimension five;
for instance,
since LRMs can be embedded in theories with extended gauge groups of associated energy scales much higher than \(v_R\), they could provide a completion for such non-renormalizable operators.
Defining \(\tilde{\chi}_{L,R}\coloneqq i\sigma^2\chi_{L,R}^{\ast}\), we have the following mass terms for quarks and leptons:

\begin{equation}
    \label{ecu.2.28}
    \mathcal{L}^q_Y=-\frac{1}{\Lambda}C_d^{ij}\bar{q}_L^{\,i}\chi_L\chi_R^{\dagger}q_R^{\,j}-\frac{1}{\Lambda}C_u^{ij}\bar{q}_L^{\,i}\tilde{\chi}_L\tilde{\chi}_R^{\dagger}q_R^{\,j}+\textrm{h.c.},
\end{equation}

\begin{equation}
    \label{ecu.2.29}
    \begin{split}
        \mathcal{L}^\ell_Y=&-\frac{1}{\Lambda}C_e^{ij}\bar{\ell}_L^{\,i}\chi_L\chi_R^{\dagger}\ell_R^{\,j}-\frac{1}{\Lambda}C_{\nu_{D}}^{ij}\bar{\ell}_L^{\,i}\tilde{\chi}_L\tilde{\chi}_R^{\dagger}\ell_R^{\,j}\\
        &-\frac{1}{\Lambda}C_{\nu_{L,M}}^{ij}\bar{\ell}_L^{\,i}\tilde{\chi}_L\tilde{\chi}_L^{\mathrm{T}}\ell_L^{j\,c}-\frac{1}{\Lambda}C_{\nu_{R,M}}^{ij}\bar{\ell}_R^{\,c\,i}\tilde{\chi}_R^{\ast}\tilde{\chi}_R^{\dagger}\ell_R^{\,j}+\textrm{h.c.},
    \end{split}
\end{equation}

\noindent with \(\Lambda\) the scale of physics beyond LRM. We have introduced a Majorana mass term for neutrinos. Indeed, the complete mass term is 

\begin{equation}
    \label{ecu.2.30}
    \mathcal{L}_m=-\sum_{f=u,\,d,\,e}\bar{f}_L\,M_f\,f_R-\frac{1}{2}\left(\bar{\nu}_L\,\bar{\nu}_R^{\,c}\right)\left(\begin{array}{cc}
     M_{\nu_{L,M}} & M_{\nu_{D}} \\
    M^{\mathrm{T}}_{\nu_{D}} & M_{\nu_{R,M}}
    \end{array}\right)\left(\begin{array}{c}
    \nu_L^{c} \\
    \nu_R
    \end{array}\right)+\textrm{h.c.},
\end{equation}

\noindent where the mass matrices are given by

 \begin{equation}
    \label{ecu.2.31}
    M_f\coloneqq \frac{1}{2}\frac{v_Lv_R}{\Lambda}C_{f},\hspace{0.4cm}M_{\nu_{D}}\coloneqq\frac{1}{2}\frac{v_Lv_R}{\Lambda}C_{\nu_{D}}, \hspace{0.4cm}M_{\nu_{L,M}}\coloneqq\frac{v_L^2}{\Lambda}C_{\nu_{L,M}},\hspace{0.4cm}M_{\nu_{R,M}}\coloneqq\frac{v_R^2}{\Lambda}C_{\nu_{R,M}}.
\end{equation}

For comparable Dirac and Majorana Wilson coefficients,
the masses of the heavy neutrinos are proportional to \(v_R^2/\Lambda\), while the masses of the light ones to \(v_L^2/\Lambda\). The masses of the other fermions are proportional to \(v_Lv_R/\Lambda\).\footnote{At tree level, the top mass is controlled by $ v_L \times (C^{(5)} \, v_R) / \Lambda $ (since $v_L$ sets the EW scale, $ \Lambda \sim C^{(5)} \, v_R $), where $ C^{(5)} $ is the Wilson coefficient of the operator of dimension five (with flavour indices omitted) and $ \Lambda \gg v_R $ is the energy scale beyond LRM. The one-loop correction goes as $ \sim v_L \times (C^{(5)} \, v_R)^3 / (16 \, \pi^2 \, \Lambda^3) $, and is thus suppressed by $1/(16 \, \pi^2)$.} The interaction between scalar fields and charged fermions is described by


\begin{equation}
    \label{ecu.2.32}
    \mathcal{L}^{Y}_{u,d,e}=-\left(1+\frac{\chi_L^{0r}}{v_L}\right)\left(1+\frac{\chi_R^{0r}}{v_R}\right)\sum_{f=u,\,d,\,e}\bar{f}\mathcal{M}_ff,
\end{equation}

\noindent  where \(\mathcal{M}_f\) is their diagonalized mass matrix. Note that in this model there are no FCNCs in the hadronic sector induced by scalar fields.
Again, since fermions come in sequential generations of identical quantum numbers, there are no FCNCs in the gauge sector at the tree level \cite{Glashow:1970gm}.


\section{Production bounds from LHC Run~2}\label{sec:direct_bounds}

\begin{table}[th!]
    \centering
    \renewcommand{\arraystretch}{1.1}
    \begin{tabular}{|lcc|c|}
        \hline
        Mode & Scenario & Eqs. & Comments \\
        \hline
        \hline
        $ Z_R \to f \bar{f} $ & All & \eqref{ecu.3.4} & \\ 
        $ Z_R \to W_L^+ W_L^- $ & All & \eqref{ecu.3.5} to \eqref{ecu.3.7} & Equivalence Theorem \\
        $ Z_R \to Z_L h^0 $ & All & \eqref{ecu.3.5} to \eqref{ecu.3.7} & Equivalence Theorem \\
        \hline
        \hline
        $ Z_R \to Z_L H^0 $ & All & \eqref{ecu.3.5} to \eqref{ecu.3.7} & ``Light new scalars'', Equiv. Theo. \\
        $ Z_R \to W_L^\pm H^\mp $ & D, T & \eqref{ecu.3.5} and \eqref{ecu.3.6} & ``Light new scalars'', Equiv. Theo. \\
        $ Z_R \to h^0 H^0 $ & All & \eqref{ecu.3.5} to \eqref{ecu.3.7} & ``Light new scalars'' \\
        $ Z_R \to W_L^\pm W_R^\mp $ & D, T & -- & Negligible \\
        \hline
        \hline
        $ Z_R \to H_1 H_2 $ & D, T & \eqref{ecu.3.5} and \eqref{ecu.3.6} & ``Light new scalars'' \\
        $ Z_R \to W_R^\pm H^\mp $ & D, T & -- & Negligible \\ 
        $ Z_R \to W_R^+ W_R^- $ & All & \eqref{ecu.3.10} and \eqref{ecu.3.11} & \\
        \hline
        \hline
        $ W_R^\pm \to f \bar{f}' $ & All & \eqref{ecu.3.15} & \\
        $ W_R^\pm \to W_L^\pm Z_L $ & D, T & \eqref{ecu.3.16} and \eqref{ecu.3.17} & Equivalence Theorem \\
        $ W_R^\pm \to W_L^\pm h^0 $ & D, T & \eqref{ecu.3.16} and \eqref{ecu.3.17} & Equivalence Theorem \\
        \hline
        \hline
        $ W_R^\pm \to W_L^\pm H^0 $ & D, T & \eqref{ecu.3.16} and \eqref{ecu.3.17} & ``Light new scalars'', Equiv. Theo. \\
        $ W_R^\pm \to H^\pm Z_L $ & D, T & \eqref{ecu.3.16} and \eqref{ecu.3.17} & ``Light new scalars'', Equiv. Theo. \\
        $ W_R^\pm \to H^\pm h^0 $ & D, T & \eqref{ecu.3.16} and \eqref{ecu.3.17} & ``Light new scalars'' \\
        $ W_R^\pm \to W_L^\mp H^{\pm\pm} $ & T & -- & Negligible \\
        \hline
        \hline
        $W_R^{\pm}\to H^{\pm} A$ & -- & -- & No contribution \\
        \hline
        \hline
        $ W_R^\pm \to H_1 H_2 $ & D, T & \eqref{ecu.3.16} and \eqref{ecu.3.17} & ``Light new scalars'' \\
        \hline
    \end{tabular}
    \caption{Set of $Z_R$ and $W_R$ two-body decays; the masses of the $Z_R$ and $W_R$ are assumed to lay much above the EW scale (which is corroborated by, e.g., the present analysis), and $ M_{W_R} \leq M_{Z_R} $ in the models analysed here: (D) two doublets and one bi-doublet, (T) two triplets and one bi-doublet, (Eff) two doublets and no bi-doublet. The masses of the RH neutrinos are discussed in the text. The notation is as follows: $ h^0, f, Z_L, W_L^\pm $ are the SM-like Higgs, fermions, neutral and charged weakly-coupled massive gauge bosons, respectively, while $A$ stands for the photon; $ Z_R, W_R^\pm $ are the new neutral and charged weakly-coupled gauge bosons, respectively; $ H $ represents the new physical scalar sector (neutral, singly or doubly charged), where we do not make explicit their $\mathcal{CP}$ nature in the case of neutral scalars (we also generally omit in this table an indexing when multiple scalars of the same charge are possible). The last column gives the ingredients used to set upper bounds on the partial widths of the $Z_R$ and $W_R$ gauge bosons, that translate into a more model-independent bound on their masses, where considering the limit of light new scalars is used only to determine the maximum of the heavy gauge-boson widths.
    }
    \label{tab:ZR_WR_two_body}
\end{table}


\subsection{Production of $Z_R$}\label{sec:direct_bounds_Zprime}



We now study direct searches involving \(Z_R\) decaying into a pair of leptons \(\ell^+\ell^-\). 
This process provides a straightforward way of putting bounds on the LR scale, since at tree level there is no dependence on the leptonic mixing matrices.
As we proceed, we will clarify that the only unknown parameters are the angle \(\gamma\) and the mass of the \(Z_R\) boson.

We study the case in which the gauge boson \(Z_R\) is produced on-shell after the collision of two protons. In particular, we want to calculate the cross section of the process \(pp\rightarrow Z_RX\rightarrow \ell^+\ell^-X\) where \(X\) represents an arbitrary state. All of the couplings needed are summarized in App.~\ref{sec:summary_couplings}. The interaction Lagrangian of the \(Z_R\) and two fermions can be written in the following two equivalent ways

\begin{equation}
    \label{ecu.3.1}
    \mathcal{L}_{Z_R}^{ff}=\frac{1}{2}Z_R^{\mu}\bar{f}\gamma_{\mu}\left(g_V^f-g_A^f\gamma_5\right)f=Z_R^{\mu}\bar{f}\gamma_{\mu}\left(g_L^f\mathcal{P}_L+g_R^f\mathcal{P}_R\right)f.
\end{equation}

\noindent In the context of \(Z_R\) searches, it is common to define the functions

\begin{equation}
    \label{ecu.3.2}
    c_q^f\coloneqq \frac{1}{2}\left( g_V^{q\,2}+g_A^{q\,2} \right)\mathrm{Br}\left(Z_R\rightarrow f\bar{f}\right)=\left(g_L^{q\,2}+g_R^{q\,2}\right)\mathrm{Br}\left(Z_R\rightarrow f\bar{f}\right) ,
\end{equation}

\noindent because for narrow resonances we have (more details will be provided in the next section when discussing $W_R^\pm$ production) \cite{Ellis:1996mzs,Workman:2022ynf}

\begin{equation}
    \label{ecu.3.3}
    \sigma\left(pp\rightarrow Z_RX\rightarrow f\bar{f}X\right)\approx\frac{\pi}{6s}\sum_qc_q^{f}\,\omega_q\!\left(s,M_{Z_R}^2\right)
\end{equation}

\noindent where \(s\) is the square of the center-of-mass energy and the \(\omega_q\!\left(s,M_{Z_R}^2\right)\) contain all the information about the Parton Distribution Functions (PDFs) of the quarks inside the proton (see the analogous Eq.~\eqref{ecu.3.14} below for the case of charged currents). The largest functions are  the ones for the up and down quarks, $q=u, d$. Since the couplings of the \(Z_R\) are generation-independent we only need to calculate \(c_u^\ell\) and \(c_d^\ell\). Note that, although $g_{V, A}^q$ or $g_{L, R}^q$ do not depend on the specific realization of the scalar sector, $c^{\ell}_{u,d}$ do.

We now calculate \(\mathrm{Br}\left(Z_R\rightarrow  \ell^+\ell^-\right)\), for which the total width of the \(Z_R\) boson \(\Gamma_{Z_R}\) is needed.
All calculations in this paper are performed at the tree level.
The possible two-body contributions are displayed in Tab.~\ref{tab:ZR_WR_two_body}.\footnote{Higher multiplicity decays are comparatively suppressed by the allowed phase space, and by powers of the weak couplings.
They cannot be enhanced
by factors of $ M_{heavy}^{n+1}/M_{light}^n \sim v_R^{n+1} / v_{\rm EW}^n $, $ n \geq 1 $, where $ M_{light} $ is the mass of an SM-like boson, since the massless limit $ M_{light} \to 0^+ $ should be continuous in a theory with SSB (for an example in which the gauge coupling of the spontaneously broken symmetry is vanishingly small, see Ref.~\cite{Weinberg:1996kr}).}
The explicit calculation of the processes that involve scalars requires diagonalizing their mass matrices.
The spectrum depends on the scalar potential, which involves a large number of parameters even when discrete symmetries are enforced.
In order to avoid this inconvenience we put an upper bound on those partial widths, which ultimately results in a model-independent bound on the mass of the \(Z_R\) boson.
A detailed discussion will follow in the coming sections.

We start by calculating the partial width into fermion pairs. It is easy to show that  if \(M_{Z_R}\gg m_f\), then

\begin{equation}
    \label{ecu.3.4}
    \Gamma\left(Z_R\rightarrow f\bar{f}\right)\approx N_C^f\frac{M_{Z_R}}{48\pi}\left(g_V^{f\,2}+g_A^{f\,2}\right)=N_C^f\frac{M_{Z_R}}{24\pi}\left(g_L^{f\,2}+g_R^{f\,2}\right)
\end{equation}

\noindent where \(N_C^f\) represents the number of colours of the fermion \(f\). Considering the mass of the fermions leads to corrections of order \((m_f/M_{Z_R})^2\), e.g., for \(M_{Z_R}\sim1\,\rm TeV\) we have for the top quark \((m_t/M_{Z_R})^2\sim 0.03\), and thus these effects are totally negligible for the SM-like fermions. However, such kinematic correction might be important for RH neutrinos
because their mass can be proportional to the LR scale. To avoid introducing new unknown parameters, we can consider the following two limiting cases: 1) very heavy RH neutrinos, where the process \(Z_R\rightarrow\nu_R\bar{\nu}_R\) is forbidden; 2) light RH neutrinos, that would give us the maximum possible value of the total width \(\Gamma_{Z_R}\) when their masses are neglected. This corresponds to the minimum value of \(\mathrm{Br}\left(Z_R\rightarrow \ell^+\ell^-\right)\), leading to less constraining bounds on the \(Z_R\) mass. Consequently, we will only study the second case in this paper.

We now consider processes that involve at least one light gauge boson in the final state.
For \(Z_R\rightarrow W_L^+W_L^-\) and \(Z_R\rightarrow Z_L h^0\) the masses of the particles in the final states are completely negligible compared to their energy, since the $Z_R$ is much heavier. In the decays \(Z_R\rightarrow  H^{\pm}W_L^{\mp}\) and \(H^0Z_L\) either the mass difference between the heavy states is small and the partial widths are suppressed by the phase space, or the mass difference is enough to allow the light particle to have a high energy compared to its own mass. Since we want to set an upper bound on the partial width of these processes, we consider the second case, in which we have light particles in the final state carrying a large momentum. Consequently, the dominant contribution to the partial widths will come from cases in which the gauge bosons in the final states have longitudinal polarization. Thus, we can calculate the amplitudes substituting the spin-1 fields by their associated Goldstone bosons using the equivalence theorem \cite{Cornwall:1974km,Lee:1977eg}, and these processes will be collectively called \(Z_R\rightarrow 2\,\mathrm{Scalars}\). In applying this theorem, we neglect the $ v_{EW}^2/v_R^2 $ corrections from the transverse degrees of freedom.

The masses of the scalar sector also determine whether the channels $Z_R \to h^0 H^0, H_1 H_2$ are allowed.
In App.~\ref{sec:Non-Fermionic_Decays} we show that we can set an upper bound on the total width of $Z_R$ using the limit where the mass of all
scalars is set to zero. Once more, this translates into the least constraining lower bounds on the mass of the \(Z_R\).
It is worth noting that when considering ``light'' scalars we avoid detailing the specific realization of the scalar sector further, i.e., whether a discrete symmetry is enforced in the scalar potential. 
Finally, we get for doublets and triplets

\begin{equation}
    \label{ecu.3.5}
    \Gamma_D\left(Z_R\rightarrow 2\,\mathrm{Scalars}\right)<\frac{M_{Z_R}}{96\pi}\frac{e^2}{\cos^2\theta_W}\left\{2\tan^2\gamma+3\cot^2\gamma\right\},
\end{equation}

\begin{equation}
    \label{ecu.3.6}
    \Gamma_T\left(Z_R\rightarrow 2\,\mathrm{Scalars}\right)<\frac{M_{Z_R}}{16\pi}\frac{e^2}{\cos^2\theta_W}\left\{\cot^2\gamma+2\tan^2\gamma\right\}.
\end{equation}

Among the previous processes, the only ones we have to consider in the Effective LR Model are \(Z_R\rightarrow  Z_Lh^0,Z_LH^0,h^0H^0\) and \(W_L^+W_L^-\). The first three decays can be taken into account by putting a bound on \(\Gamma_{\mathrm{Eff}}\left(Z_R\rightarrow2\,\mathrm{Neutral\,Scalars}\right)\) as we did for the other two models. We can compute exactly the partial width of \(Z_R\rightarrow W_L^+W_L^-\) using the Feynman rule in Eq.~\eqref{ecu.A.9}:
 

\begin{equation}\label{ecu.3.7}
    \begin{split}
    & \Gamma_{\mathrm{Eff}}\left(Z_R\rightarrow 2\,\mathrm{Neutral\,Scalars}\right)+\Gamma_{\mathrm{Eff}} \left(Z_R\rightarrow W_L^+W_L^-\right)
    \\& \;\;\;\;\;\;\; <\frac{M_{Z_R}}{192\pi}\frac{e^2}{\cos^2\theta_W}\left\{2+\cot^2\gamma+3\tan^2\gamma\right\}.
    \end{split}
\end{equation}
 
 We still need to study the decays \(Z_R\rightarrow W_R^{\pm}W_L^{\mp},\,H^{\pm}W_R^{\mp}\) and \(W_R^+W_R^-\). In App.~\ref{sec:Non-Fermionic_Decays} we show that the contribution of the first two is negligible since the partial width is, at most, proportional to \(v_{EW}^2/v_R\).
%
%
Finally, the exact value of the partial width for the process \(Z_R\rightarrow W_R^+W_R^-\) can be written as

\begin{equation}
    \label{ecu.3.10}
    \Gamma_D\left(Z_R\rightarrow W_R^+W_R^-\right)=\frac{M_{Z_R}}{192\pi}\left(\frac{e}{\cos\theta_W}\right)^2\!\! \; \frac{\left(1-4\cos^2\gamma\right)^{3/2}}{\sin^2\gamma\cos^2\gamma}\!\left\{1+20\cos^2\gamma+12\cos^4\gamma\right\},
\end{equation}

\begin{equation}
    \label{ecu.3.11}
    \Gamma_T\left(Z_R\rightarrow W_R^+W_R^-\right)=\frac{M_{Z_R}}{48\pi}\left(\frac{e}{\cos\theta_W}\right)^2\!\! \; \frac{\left(1-2\cos^2\gamma\right)^{3/2}}{\sin^2\gamma\cos^2\gamma}\!\left\{1+10\cos^2\gamma+3\cos^4\gamma\right\}.
\end{equation}

As evident, Eq.~(\ref{ecu.3.10}) also applies to the effective LR model with only doublet scalars.
This process gives a result only if \(\gamma\in\left[60^{\circ},70^{\circ}\right]\) or \(\gamma\in\left[45^{\circ},70^{\circ}\right]\) in the models with doublets or triplets, respectively, given the requirement $ M_{Z_R} > 2 \, M_{W_R} $, and that we stay in the perturbative regime (\(70^{\circ}\gtrsim\gamma\gtrsim20^{\circ}\)). 

\begin{figure}
    \centering
    \includegraphics[scale=0.7]{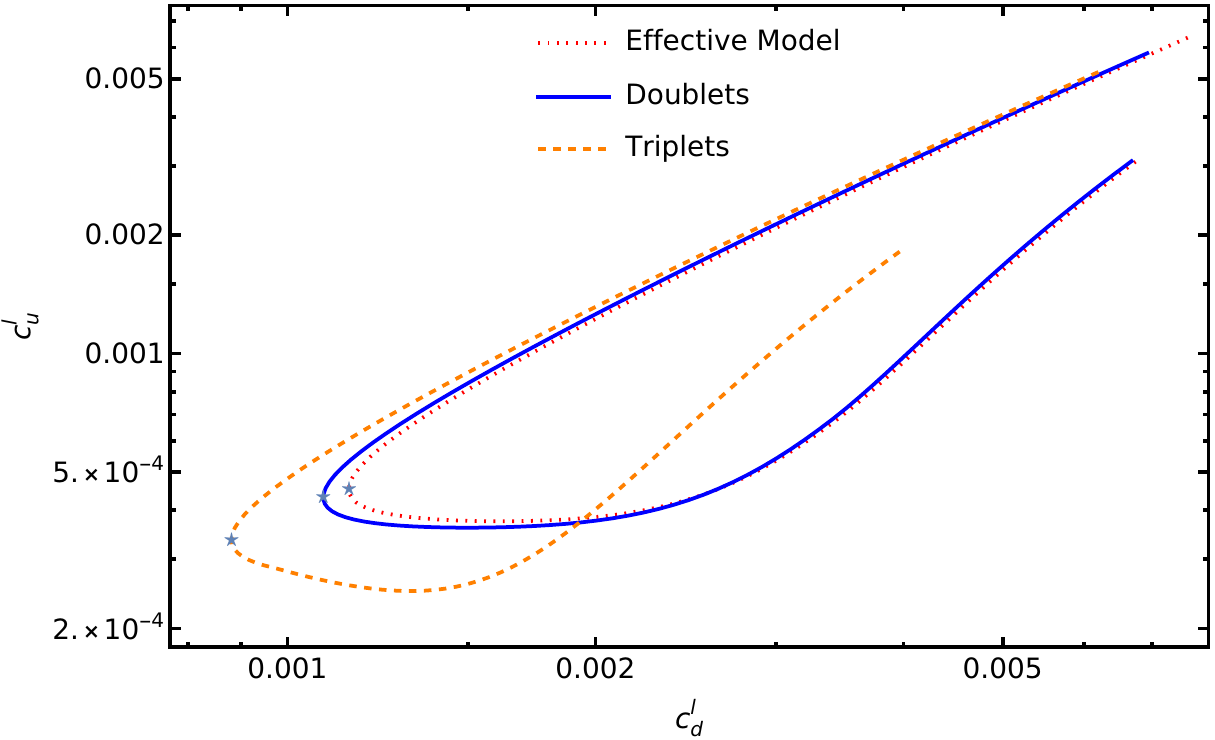}
    \caption{Parametric plot of the functions \(c^\ell_d\) and \(c^\ell_u\) depending on the angle \(\gamma\). Varying $\gamma$ in the range $ [20^{\circ}, 70^{\circ}] $, the plotted curves are transited counterclockwise. The points used for setting the bounds on the mass of $Z_R$ are the ones marked with a star, see Fig.~\ref{fig:Gamma_doublets_triplets}. They correspond to values of \(\gamma\) around \(40^{\circ}\) and \(41^{\circ}\).}
    \label{fig:cu_cd_plane}
\end{figure}

\begin{figure}
    \centering
    \includegraphics[scale=0.67]{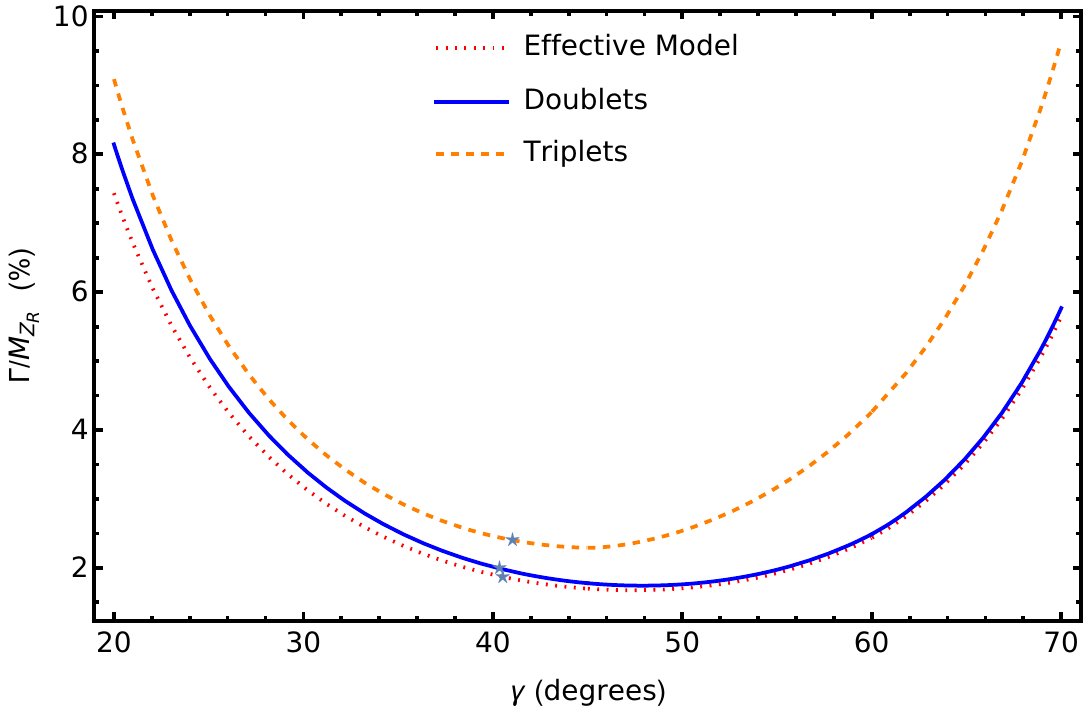}
    \caption{Total width of the $Z_R$ as a function of the angle $\gamma$. The points marked with a star are the ones used for setting the bounds on the mass $M_{Z_R}$, see Fig.~\ref{fig:cu_cd_plane}.}
    \label{fig:Gamma_doublets_triplets}
\end{figure}

We observe that all 
partial widths are proportional to \(M_{Z_R}\), with no further explicit dependence on this quantity. Thus, both \(c_u^\ell\) and \(c_d^\ell\) will only depend on \(\gamma\). 
This dependence is displayed in Fig.~\ref{fig:cu_cd_plane} for the three LR models considered. The lowest cross sections are obtained when $c_d^\ell$ reaches its minimum possible value allowed by this model; the results are very close to the minimum of $c_u^\ell$, which shows a flatter dependence on $\gamma$ compared to $c_d^\ell$.
Thus, when comparing to the experimental data, the three points marked with a star in Fig.~\ref{fig:cu_cd_plane} will give the less constraining lower bound on the $Z_R$ mass, which in
turn are close to the LR limit $g_L = g_R$.
For the values of \(\gamma\) used to set bounds on the mass of $Z_R$, the values of the total width over the mass of the \(Z_R\) gauge boson are below \(3\%\), as shown in Fig.~\ref{fig:Gamma_doublets_triplets}. Using the narrow-resonance 
approximation is thus justified.

The current experimental information can be extracted from the following LHC
searches for two charged leptons in the final state:

\begin{itemize}
    \item CMS ($ \sqrt{s} = 13 $~TeV) \cite{CMS:2021ctt}, in the di-electron (137~fb$^{-1}$) and di-muon (140~fb$^{-1}$) channels;
    \item ATLAS ($ \sqrt{s} = 13 $~TeV, 139~fb$^{-1}$) \cite{ATLAS:2019erb}, in the di-electron and di-muon channels;
    \item ATLAS ($ \sqrt{s} = 13 $~TeV, 36.1~fb$^{-1}$) \cite{ATLAS:2017eiz}, in the di-tau channel.
\end{itemize}

They achieve very similar bounds on the $Z_R$ mass in the former two cases (i.e., di-electron and di-muon searches), which are better than in the third case (i.e., di-tau searches, of lower statistics); we will not attempt at combining results from ATLAS and CMS in this article.
We observe a modest impact of the total width of the $Z_R$ on the bounds on its mass in the experimental searches of Refs.~\cite{CMS:2021ctt,ATLAS:2019erb}, for values of the mass $ M_{Z_R} \gtrsim 4 $~TeV, and when $ \Gamma_{Z_R} \lesssim 10 \% $.
Theoretical uncertainties result from PDF uncertainties and higher-orders, and grow with the invariant mass, reaching about 20\% at scales of many TeV.
Based on Ref.~\cite{CMS:2021ctt} for the combination of di-electron and di-muon modes, we achieve the bounds shown in Tab.~\ref{tab:summary_bounds}.

There are also available
di-jet searches from CMS ($ \sqrt{s} = 13 $~TeV, 137~fb$^{-1}$) \cite{CMS:2019gwf} and ATLAS ($ \sqrt{s} = 13 $~TeV, 139~fb$^{-1}$) \cite{ATLAS:2019fgd},
and searches based on the di-boson decay modes $ Z_R \to W_L^+ W_L^- $ and $ Z_R \to Z_L h^0 $, Refs.~\cite{ATLAS:2019nat,ATLAS:2020fry,CMS:2021fyk,CMS:2021klu}.
Better bounds, however, are achieved from the previous di-lepton searches, see also Refs.~\cite{Osland:2020onj,Osland:2022ryb}.

Our very conservative lower bounds on the \(Z_R\) mass remain valid even if we relax the condition \(g_R,\,g_X\leq 1\) because, for values of \(\gamma\) smaller than \(20^{\circ}\) or greater than \(70^0\), we would get larger values for both \(c_d^\ell\) and \(c_u^\ell\), as shown in Fig.~\ref{fig:cu_cd_plane}. Of course, with larger predicted cross sections one could get stronger bounds, but they would be model dependent. Instead, we are adopting the more pessimistic scenario (minimum values of $c_d^\ell$) in order to obtain a robust model-independent lower limit on $M_{Z_R}$.

\subsection{Production of $W_R$}\label{sec:direct_bounds_Wprime}

We now discuss searches involving the production and decay to a pair of fermions of the \(W_R\) boson in colliders. In this case we have more free parameters because the elements of the mixing matrix \(V_R^{\mathrm{CKM}}\) appear explicitly. We want to calculate the cross section of the process \(pp\rightarrow W_R^\pm X \rightarrow f\bar{f}' X\).
The general expression
in the s-channel and at leading order in \(\alpha_s\)
is the following \cite{Ellis:1996mzs}:

\begin{eqnarray}\label{eq:beyond_NWA_1}
    \lefteqn{\sigma\left(pp\rightarrow W_R^\pm X \rightarrow f\bar{f}' X\right) = \frac{2\pi}{3} \frac{\Gamma_{W_R}}{M_{W_R}} \alpha_R\sum_{ij}|\!\left(V_R^{\mathrm{CKM}}\right)_{ij}\!|^2} && \\
    & \times & \int_{0}^1 {\rm d}x_1 {\rm d}x_2 \left[u_i\left(x_1,\,\mu\right)\bar{d}_j\left(x_2,\,\mu\right)+\bar{u}_i\left(x_2,\,\mu\right)d_j\left(x_1,\,\mu\right)\right] \frac{\hat{s} \, \mathrm{Br}\left(W_R^\pm \rightarrow f\bar{f}'\right)}{(\hat{s} - M^2_{W_R})^2 + M^2_{W_R} \, \Gamma^2_{W_R}}\, , \nonumber
\end{eqnarray}
where $ \hat{s} \coloneqq x_1 x_2 s $.
%
%
For narrow resonances, one can substitute\footnote{Equivalently, for numerical purposes one can take 
$ \hat{s} $ in the integration range $ [(M_{W_R} - \Delta)^2, (M_{W_R} + \Delta)^2] $ (together with requiring $ \hat{s} \leq x s $, $ x $ being the other integration variable), where $ \Gamma_{W_R} \ll \Delta \ll M_{W_R} $.}

\begin{equation}
    \frac{1}{(\hat{s} - M^2_{W_R})^2 + M^2_{W_R} \, \Gamma^2_{W_R}} \;\to\; \frac{\pi}{M_{W_R} \, \Gamma_{W_R}} \,\delta (\hat{s} - M^2_{W_R})\, ,
\end{equation}
leading to:

\begin{equation}
    \label{ecu.3.12}
    \sigma\left(pp\rightarrow W_R^\pm X \rightarrow f\bar{f}' X\right)\approx\sigma\left(pp\rightarrow W_R^\pm X \right)\mathrm{Br}\left(W_R^\pm \rightarrow f\bar{f}'\right)
\end{equation}


\noindent after integration over $ \hat{s} $. The $W_R^\pm$ production cross section is \cite{Workman:2022ynf}

\begin{equation}
    \label{ecu.3.13}
    \sigma\left(pp\rightarrow W_R^\pm X \right)\approx\frac{2\pi^2}{3s}\alpha_R\sum_{ij}|\!\left(V_R^{\mathrm{CKM}}\right)_{ij}\!|^2\,\omega_{ij}\!\left(M^2_{W_R}/s,\,M_{W_R}\right),
\end{equation}

\noindent where \(s\) is the square of the center-of-mass energy, \(\alpha_R\coloneqq g_R^2/(4\pi)\) and the functions \(\omega_{ij}\left(z,\mu\right)\) are given by

\begin{equation}
    \label{ecu.3.14}
    \omega_{ij}\left(z,\,\mu\right)=\int_{z}^1\frac{{\rm d}x}{x} \left[u_i\left(x,\,\mu\right)\bar{d}_j\left(z/x,\,\mu\right)+\bar{u}_i\left(x,\,\mu\right)d_j\left(z/x,\,\mu\right)\right],
\end{equation}

\noindent with \(q_i\left(x,\,\mu\right)\) the PDF of the quark \(q_i\) inside the proton at the factorization scale \(\mu\) and parton momentum fraction \(x\) (sub-leading effects in \(\alpha_s\) will be discussed ahead).
Note that there is an explicit dependence on the elements of the matrix \(V_R^{\mathrm{CKM}}\) in Eq.~(\ref{ecu.3.13}).
Feynman rules for the $W_R$ couplings to fermion pairs are found in Eqs.~(\ref{ecu.A.11}) and (\ref{ecu.A.12}).

As derived from Eq.~\eqref{eq:beyond_NWA_1}, not restricting the $W_R$ from being produced with an invariant mass close to its resonance peak
largely enhances the total cross section for large values of $ M_{W_R} $, compared to the narrow-width approximation, when $ M_{W_R} $ approaches $ \sqrt{s} $: in this case $\omega_{ij}$ in Eq.~\eqref{ecu.3.14} approaches $0$ since $z$ approaches $1$, while non-vanishing values are achieved from the original expression Eq.~\eqref{eq:beyond_NWA_1} when not restricting the value of $\hat{s}$.
At the same time, the SM background increases when allowing for small values of $\hat{s}$, so one loses out on sensitivity towards signal events.


We now calculate the decay width of the \(W_R\) boson \(\Gamma_{W_R}\), at leading order in \(g_R\). At tree level, the relevant processes are listed in Tab.~\ref{tab:ZR_WR_two_body}.
Considering that the fermions in the final state have a mass much smaller than the one of the \(W_R\) boson it is easy to obtain

\begin{equation}
    \label{ecu.3.15}
    \Gamma\left(W_R\rightarrow f_i\bar{f}_j\right)\approx\frac{\alpha_R}{12}|\!\left(V_R\right)_{ij}\!|^2N^f_CM_{W_R},
\end{equation}

\noindent where \(N^f_C\) represents the number of colours of the fermion $f$ and \(V_R=V_R^{\mathrm{CKM}}\) or \(V_R=V_R^{\mathrm{PMNS}}\) for quarks and 
leptons, respectively. In order to avoid the introduction of new free parameters, we can consider the two limiting cases for the masses of the RH neutrinos. If they are extremely heavy only the quark decay modes are kinematically open and
we have \(\Gamma\left(W_R\rightarrow f\bar{f}'\right) \coloneqq \sum_{ij} \Gamma\left(W_R\rightarrow f_i\bar{f}_j\right)
\approx (3/4)\alpha_RM_{W_R}\), while for light RH neutrinos we get \(\Gamma\left(W_R\rightarrow f\bar{f}'\right)\approx\alpha_RM_{W_R}\). The less stringent bound on the mass of the \(W_R\) results from the second case.

When using the equivalence theorem \cite{Cornwall:1974km,Lee:1977eg}, it is easy to see that we can set a bound on the sum of partial widths of the decays \(W_R^{\pm}\rightarrow W_L^{\pm}Z_L\), $ W_L^{\pm}h^0 $, $ W_L^{\pm}H^0 $, $H^{\pm}Z_L$, $H^{\pm}h^0$, $H^{\pm}H^0$ and \(H^{\pm\pm}H^{\mp}\), by calculating a bound on \(\Gamma\left(W_R\rightarrow 2 \, \mathrm{Scalars}\right)\). The contribution of the process \(W_R^{\pm}\rightarrow W_L^{\mp}H^{\pm\pm}\) for the models with triplets is negligible. The decay \(W_R^{\pm}\to H^{\pm}A\) is forbidden at the tree level. The details are given in App.~\ref{sec:Non-Fermionic_Decays}.
By adding all fermionic and bosonic final states, we get
the following bounds on the total width of the \(W_R\) boson 
for the doublet and triplet models:

\begin{equation}
    \label{ecu.3.16}
    \Gamma_{W_R}^D< \frac{9}{8}\alpha_R M_{W_R},
\end{equation}

\begin{equation}
    \label{ecu.3.17}
    \Gamma_{W_R}^T< \frac{5}{4}\alpha_R M_{W_R}.
\end{equation}


The non-fermionic processes \(W_R^{\pm}\rightarrow W_L^{\pm} Z_L,W_L^{\pm}h^0\) and \(W_L^{\pm}H^0\)
are not allowed in the Effective LR Model because there is no mixing between the \(W_L\) and the \(W_R\) gauge bosons.
Thus, we obtain the following bound by considering light RH neutrinos:

\begin{equation}
    \label{ecu.18}
     \Gamma_{W_R}^{\mathrm{Eff}}< \alpha_R M_{W_R}.
\end{equation}

\begin{figure}[t]
\centering
\includegraphics[scale=0.4]{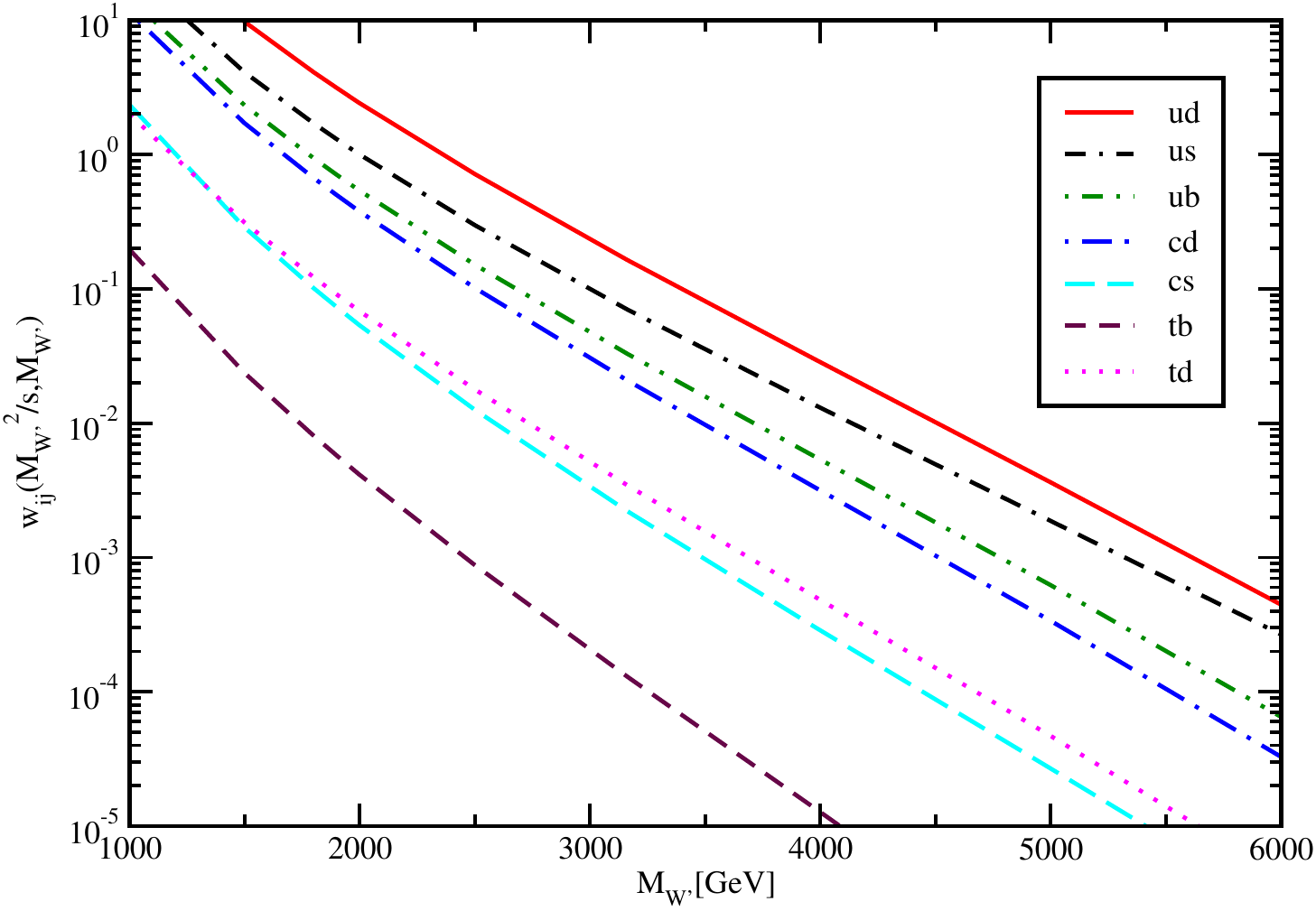}
\caption{The quantity $w_{ij}(M_{W_R}^2/s,M_{W_R})$ for up and down type quarks of various generations as a function of the $W' \coloneqq W_R$ mass. Figure extracted from Ref.~\cite{Bernard:2020cyi}.}
\label{fig:pdfatlas}
\end{figure}

The recasted bound on the mass of the \(W_R\) boson obtained from direct searches in colliders depends on the specific structure of the \(V_R^{\mathrm{CKM}}\) matrix.
For instance, we can find the values of the matrix elements  \(|\left(V_R^{\mathrm{CKM}}\right)_{ij}|\) that minimize the cross section in Eq.~(\ref{ecu.3.13}).
The functions \(\omega_{ij}\) are displayed in Fig.~\ref{fig:pdfatlas}, and have been obtained using the NNPDF23LO PDF set \cite{Ball:2012cx} (at a factorization scale of $ 6 $~TeV, and for $ \sqrt{s} = 13 $~TeV).
The figure clearly shows
that the main contributions come from \(\omega_{ud}\) and \(\omega_{us}\); then, we take \(\left(V_R^{\mathrm{CKM}}\right)_{ud}=\left(V_R^{\mathrm{CKM}}\right)_{us}=0\). After the previous two terms, the one proportional to \(\omega_{ub}\) is dominant; in this case we cannot take \(\left(V_R^{\mathrm{CKM}}\right)_{ub}=0\) as the matrix $V_R^{\mathrm{CKM}}$ must be unitary. The following leading term in the $W_R$ production is the one proportional to \(\omega_{cd}\), for which we then can set \(\left(V_R^{\mathrm{CKM}}\right)_{cd}=0\). Taking the above into account we arrive at the following structure:

\begin{equation}
    \label{ecu.3.19}
    V_R^{\mathrm{CKM}}=\left(
    \begin{array}{ccc}
    0 & 0 & 1 \\
    0 & 1 & 0 \\
    1 & 0 & 0 
    \end{array}\right)
\end{equation}
where complex phases are irrelevant for our purposes.
This anti-diagonal matrix is associated with the minimum possible value of the cross section in the s-channel, and we get from it the less stringent bounds on \(M_{W_R}\) from di-jet and di-lepton searches that we discuss in the following sections (see Figs.~\ref{fig:WR_Dijet} and \ref{fig:WR_Lepton} and Tab.~\ref{tab:summary_bounds}).
These bounds can be somewhat improved by considering NLO corrections in QCD, namely, associated production of the $W_R$ with a bottom quark. This is phase-space suppressed, and moreover suppressed by the small value of the strong coupling at the high perturbative scales relevant for the process, but depends on the PDF of the gluon.\footnote{
The $(V^{\rm CKM}_R)_{cb}$ coupling has been studied in the context of $B$-meson flavour-changing charged-current anomalies,
and it was noticed that with a future LHC luminosity of 300~fb$^{-1}$ one can probe $ \mathcal{O}(1) $ couplings of a $W_R$ of mass around $ 2.4 $~TeV in the tau decay channel,
see Ref.~\cite{Altmannshofer:2017poe}; see also Refs.~\cite{Abdullah:2018ets,Blanke:2022pjy}.}
For comparison
we also provide in Figs.~\ref{fig:WR_Dijet} and \ref{fig:WR_Lepton}, and Tab.~\ref{tab:summary_bounds} 
the stronger limits that would be obtained with
a diagonal structure 
of the RH quark mixing matrix, for which the latter associated $W_R$ production, like the t-channel production, is negligible compared to the dominant s-channel production \cite{Sullivan:2002jt}.


\subsubsection{Di-jet decay mode}


We now discuss in details the di-jet decay mode of the \(W_R\) boson.
This channel has the advantage of not depending on the characteristics of the leptonic sector, namely, neutrino masses and RH lepton mixing matrix.
In order to obtain the cross section of the process \(pp\rightarrow W_R\rightarrow j_1j_2\) we need to calculate \(\Gamma\left(W_R\rightarrow q_1\bar{q}_2\right)\). Looking at Eq.~(\ref{ecu.3.15}), we get

\begin{equation}
    \label{ecu.3.20}
    \Gamma\left(W_R\rightarrow q_1\bar{q}_2\right)\approx\frac{\alpha_R}{4}M_{W_R}\sum_{i=u,c,t}\sum_{j=d,s,b}\left|\left(V_R^{\mathrm{CKM}}\right)_{ij}\right|^2= \frac{3}{4}\alpha_RM_{W_R}.
\end{equation}

\noindent Consequently, we obtain the following lower bounds on the branching fractions:

\begin{equation}
    \label{ecu.3.21}
    \mathrm{Br}\left(W_R\rightarrow j_1j_2\right)_{\mathrm{Eff}}> \frac{3}{4},
\end{equation}

\begin{equation}
    \label{ecu.3.22}
    \mathrm{Br}\left(W_R\rightarrow j_1j_2\right)_D>\frac{2}{3},
\end{equation}

\begin{equation}
    \label{ecu.3.23}
    \mathrm{Br}\left(W_R\rightarrow j_1j_2\right)_T> \frac{3}{5}.
\end{equation}
In the hadronic $W_R^\pm$ decay width we have neglected the QCD correction factor $\delta_{\mathrm{QCD}}\approx 1 + \frac{\alpha_s}{\pi}\ge 1$, which is small at the high scale $\mu=M_{W_R}$; since it would increase the branching fractions, these lower bounds are even more conservative.

For a given mixing matrix $V_R^{\mathrm{CKM}}$, the minimum value of the production cross section corresponds to the minimum value of \(\alpha_R = \alpha/(\sin^2{\gamma} \cos^2{\theta_W}) \), 
for which we have \(\Gamma_{W_R}^D/M_{W_R}< 1.2\%\), \(\Gamma_{W_R}^T/M_{W_R}< 1.3\%\) and \(\Gamma_{W_R}^{\mathrm{Eff}}/M_{W_R}< 1.1\%\), clearly in the narrow resonance regime.
There are data from:

\begin{itemize}
    \item CMS ($ \sqrt{s} = 13 $~TeV, 139~fb$^{-1}$) \cite{CMS:2019gwf,CMS_supplementary_material_dijet} and
    \item ATLAS ($ \sqrt{s} = 13 $~TeV, 139~fb$^{-1}$) \cite{ATLAS:2019fgd,ATLAS_supplementary_material_dijet}.
\end{itemize}

\begin{figure}[t]
\centering
\includegraphics[scale=0.6]{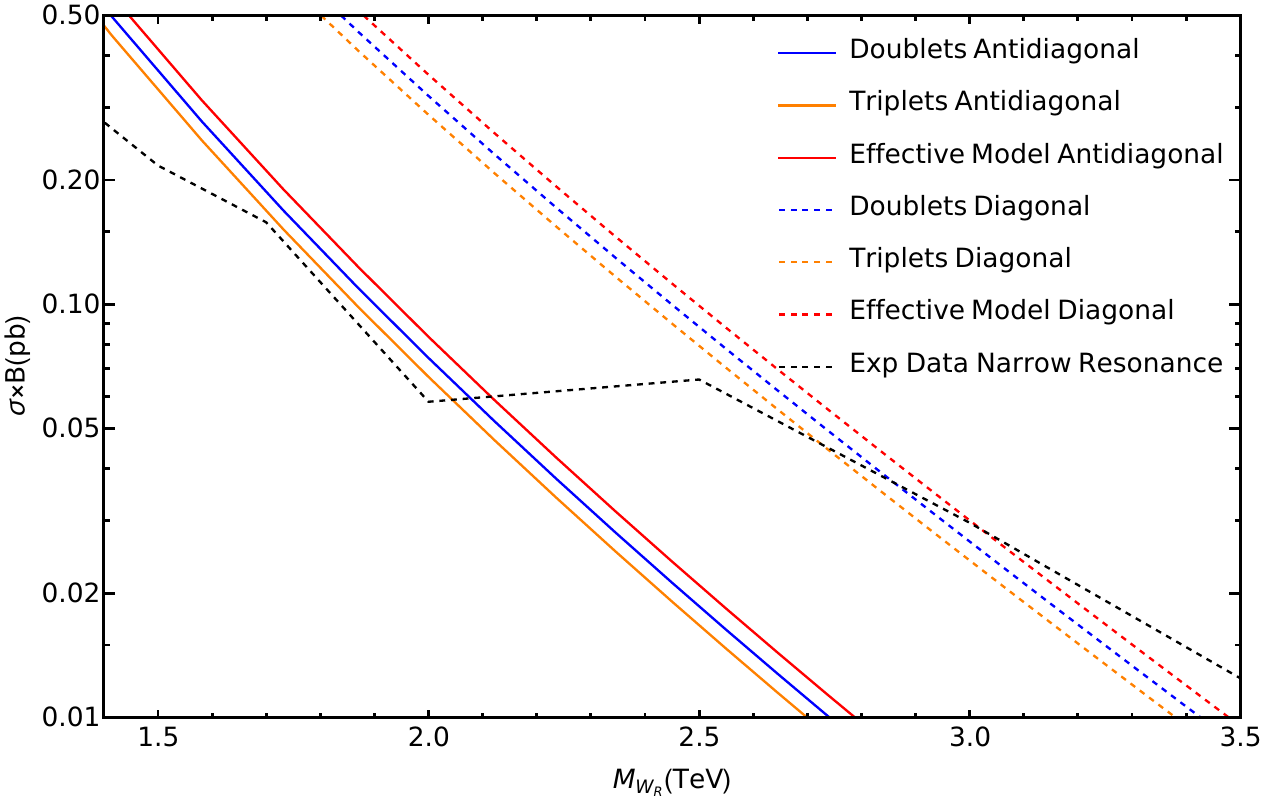}
\caption{Comparison between the theoretical prediction of the cross section times the branching fraction of the process \(W_R\rightarrow jj\) for different LR symmetric models and an anti-diagonal or diagonal structure of $V_R^{\mathrm{CKM}}$. Also shown is
the experimental bound obtained by ATLAS \cite{ATLAS:2019fgd} in the narrow resonance regime.}
\label{fig:WR_Dijet}
\end{figure}

They achieve very similar bounds on the total cross section up to $ \lesssim 4$~TeV; as previously stated, we will not attempt at combining results from ATLAS and CMS in this article.
Note that in Eq.~\eqref{ecu.3.20} we have included the top flavour, which is highly boosted and explicitly included in the simulation of signal events in Ref.~\cite{ATLAS:2019fgd}, but it is also present to some extent in the final di-jet mode of Ref.~\cite{CMS:2019gwf,andreas-hinzmann}.
The results are shown in Fig.~\ref{fig:WR_Dijet} and Tab.~\ref{tab:summary_bounds} for Ref.~\cite{ATLAS:2019fgd}, where the acceptance from Ref.~\cite{ATLAS_supplementary_material_dijet} for the $W'$ model has been taken into account.
In the comparison to data, we employ a $K$ factor of 1.3 \cite{CMS:2019gwf,Sullivan:2002jt}.

\subsubsection{Leptonic decay mode}


If the masses of the RH neutrinos are negligible compared to the mass of the \(W_R\) we can also consider the leptonic decay modes \(W_R\rightarrow \ell_i \bar{\nu}_R\), $ \ell_i = e, \mu, \tau $, where the RH neutrinos are stable (we omit their flavours). In this case, adding the contributions from the different neutrinos, one gets
\(\Gamma\left(W_R\rightarrow \ell_i \bar{\nu}_R\right)\approx\alpha_R M_{W_R}/12\). Therefore, the bounds on the branching fractions are

\begin{equation}
    \label{ecu.3.24}
    \mathrm{Br}\left(W_R\rightarrow \ell_i\bar{\nu}_R\right)_{\mathrm{Eff}}\gtrsim\frac{1}{12},
\end{equation}

\begin{equation}
    \label{ecu.3.25}
    \mathrm{Br}\left(W_R\rightarrow \ell_i\bar{\nu}_R\right)_D>\frac{2}{27},
\end{equation}

\begin{equation}
    \label{ecu.3.26}
    \mathrm{Br}\left(W_R\rightarrow \ell_i\bar{\nu}_R\right)_T>\frac{1}{15}.
\end{equation}

\begin{figure}[t]
\centering
\includegraphics[scale=0.77]{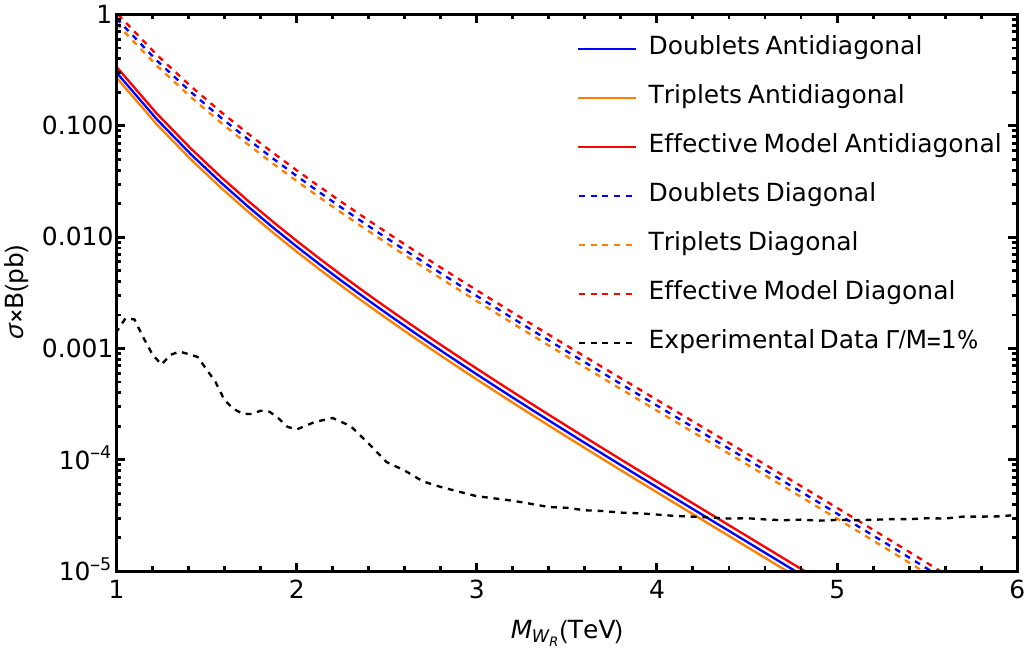}
\caption{Comparison between the theoretical prediction of the cross section times the branching fraction of the process \(W_R\rightarrow \ell_i \bar{\nu}_R\) for different LR symmetric models and an anti-diagonal or diagonal structure of $V_R^{\mathrm{CKM}}$. Also shown is the experimental bound obtained by ATLAS \cite{ATLAS:2019lsy}, combining the data from the decays to \(e \bar{\nu}_R\) and \(\mu \bar{\nu}_R\) in the narrow resonance regime. The data corresponds to the case $ m_{\ell \nu} > 0.3 \, M_{W_R}$.}
\label{fig:WR_Lepton}
\end{figure}

\noindent From the experimental side,
there are data on light lepton modes:

\begin{itemize}
    \item ATLAS ($ \sqrt{s} = 13 $~TeV, 139~fb$^{-1}$) \cite{ATLAS:2019lsy,ATLAS_supplementary_material_ellnu} and
    \item CMS ($ \sqrt{s} = 13 $~TeV, 138~fb$^{-1}$) \cite{CMS:2022yjm,CMS_supplementary_material_ellnu},
\end{itemize}
and the tauonic mode:

\begin{itemize}
    \item ATLAS ($ \sqrt{s} = 13 $~TeV, 139~fb$^{-1}$) \cite{ATLAS:2021bjk} and
    \item CMS ($ \sqrt{s} = 13 $~TeV, 35.9~fb$^{-1}$) \cite{CMS:2018fza}.
\end{itemize}

They achieve very similar bounds on the $W_R$ mass in the combined light lepton case, which are better than those bounds in the latter tauonic mode.
We take the combined data of \(W_R\rightarrow e \bar{\nu}_R\) and \(W_R\rightarrow \mu \bar{\nu}_R\) from Ref.~\cite{ATLAS:2019lsy}.
Theoretical uncertainties from PDFs and higher-orders are small.
We employ a constant $K$ factor of 1.3, valid for $ M_{W_R} \simeq 3.0 $~TeV \cite{CMS:2022yjm}.
The results of the bound on \(M_{W_R}\) are shown in Fig.~\ref{fig:WR_Lepton}. We employ the experimental data of the fiducial case $ m_{\ell \nu} > 0.3 \, M_{W_R}$, which is very close to the on-shell limit.
The less constraining bound is achieved for the case with the minimum value of \(\alpha_R\), which is shown therein.
These bounds are better than the ones produced by searches of the $ W_R^\pm \to W_L^\pm h^0 $ (which depends on the realization of the scalar sector) and $ W_R^\pm \to W_L^\pm Z_L $ decay modes, see Ref.~\cite{Workman:2022ynf} and references therein; note that both channels are not present in the Effective LR Model.

Relaxing the condition \(g_R,\, g_X\leq 1\) in the direct searches of the \(W_R\) gauge boson has an impact in the bound on its mass. This is due to the fact that, if we allow \(g_X\) to take greater values, then \(g_R\) could be smaller and consequently we would get a smaller cross section.





\begin{figure}
    \centering
    \includegraphics[scale=0.463]{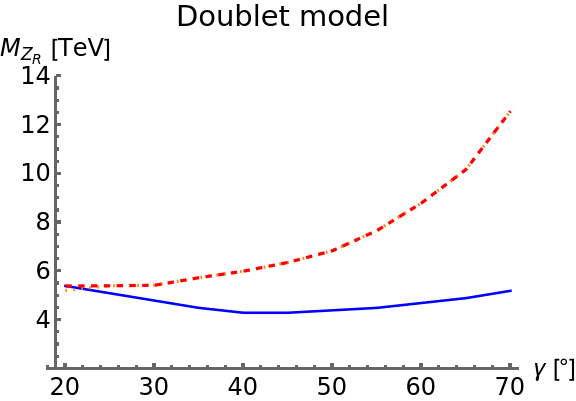} \hspace{5mm}
    \includegraphics[scale=0.463]{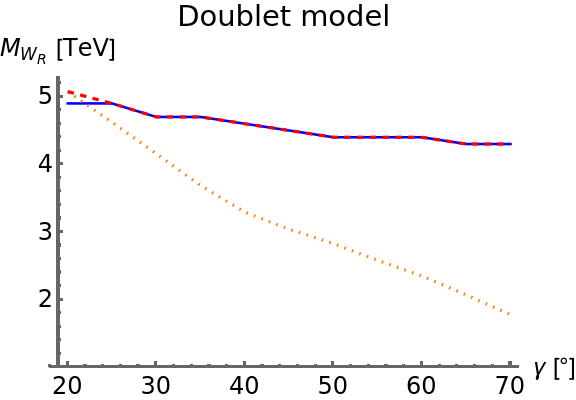} \\
    \vspace{5mm}
    \includegraphics[scale=0.37]{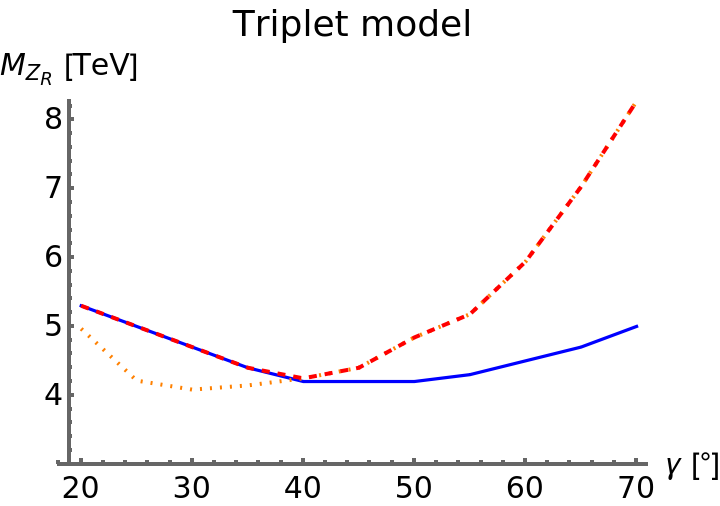} \hspace{5mm}
    \includegraphics[scale=0.37]{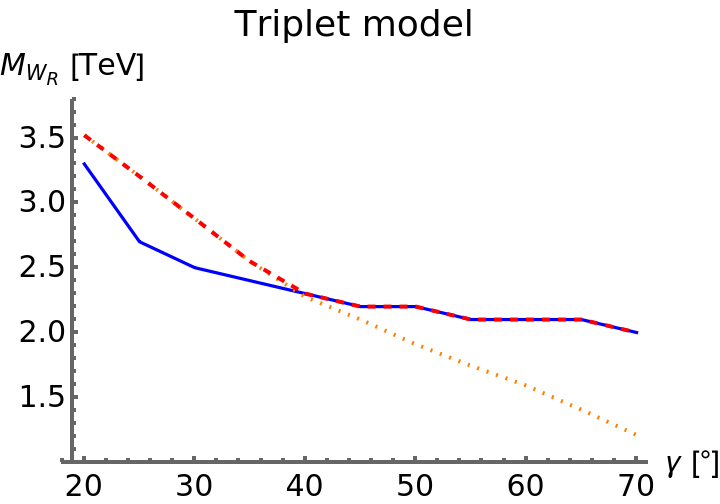} \\
    \vspace{5mm}
    \includegraphics[scale=0.463]{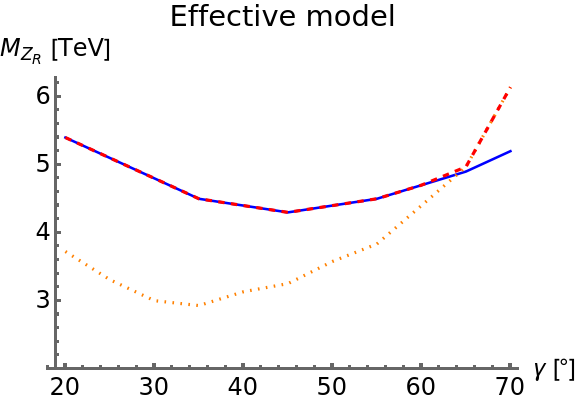} \hspace{5mm}
    \includegraphics[scale=0.463]{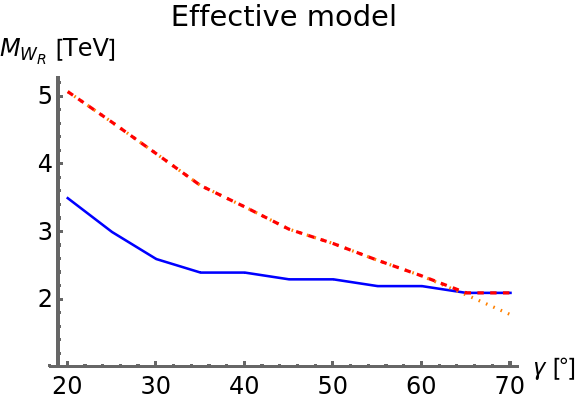}
    \caption{Bounds on 
    $M_{Z_R}$ (left panels) and $M_{W_R}$ (right panels) in
    the doublet (upper panels), triplet (middle panels) and effective models (lower panels).
    The solid blue curves correspond to the bounds on $M_{Z_R}$ ($M_{W_R}$)
    derived from direct searches of the $Z_R$ (respectively, $W_R$) gauge boson; 
    the dotted orange lines show the bounds on the $Z_R$ ($W_R$) mass derived from direct searches of the $W_R$ (respectively, $Z_R$) gauge boson, based on the relation between the two heavy gauge boson masses in the LRM realizations under discussion, see Eqs.~\eqref{eq:relation_masses_doublets} and \eqref{eq:relation_masses_triplets}.
    The strongest limits are displayed in dashed red.
    }
    \label{fig:direct_indirect_mass_bounds}
\end{figure}

\subsection{Direct and indirect bounds on $ M_{Z_R} $ and $ M_{W_R} $}\label{sec:mass_relation}



Hereby we summarize the bounds achieved in the previous sections.
They are displayed with solid blue curves in Fig.~\ref{fig:direct_indirect_mass_bounds} as function of the mixing angle $\gamma$ for the three scalar sectors analysed: (D), (T) and (Eff). The left and right panels show, respectively, the lower limits on $M_{Z_R}$ and $M_{W_R}$. 
The $W_R$ figures correspond to an anti-diagonal RH quark-mixing matrix that imposes less restrictive constraints.
In the triplet and effective cases, where Majorana neutrinos are possible, they are based on data from the $ W_R \to jj $ mode, which is independent of the neutrino sector.
In the doublet case, where neutrinos are purely Dirac fermions, the blue curve shows the more constraining bounds on $ M_{W_R} $ obtained
from the leptonic decay mode $ W_R \to \ell \bar{\nu}_R $ with light RH neutrinos.

The values in Tab.~\ref{tab:summary_bounds} result from the less stringent constraints in scenarios (D), (T) and (Eff), i.e., they correspond to the value of $\gamma$ giving the weakest lower bound in the given scenario, assuming a perturbative theory. Therefore,
masses are excluded below these values (with a high significance, of around 95\% CL).
We display in all three cases the limits on $M_{W_R}$ obtained from the hadronic and leptonic decay modes of the $W_R$, and with diagonal and anti-diagonal structures of the RH quark mixing matrix.
%
The masses are constrained to lie above a few TeV in both cases.
The most important bound on the $ W_R $ mass comes from the leptonic decay mode, which depends on the assumption that RH neutrinos are light enough. The bounds on the $ W_R $ mass from di-jet final states are independent of this assumption.
Note that
the consideration of different structures of the mixing matrix for RH quarks leads to differences of $ \lesssim 1 $~TeV in the recasted bounds.
The bounds are very similar across the three LRMs that we discuss in this work, which differ in their scalar content, implying that collider searches based on fermions in the final state are mostly independent of the impact of the scalar sector on the heavy gauge-boson total width.




\begin{table}[h]
    \centering
    \renewcommand{\arraystretch}{1.2}
    \begin{tabular}{|lcc|c|c|c|}
        \hline
        Channel & Refs. & & $\Phi$ + $\chi_{L,R}$ (D) & $\Phi$ + $\Delta_{L,R}$ (T) & $\chi_{L,R}$ (Eff) \\
        \hline
        \hline
        $Z_R \to \ell_i \bar{\ell}_i$ & \cite{CMS:2021ctt} & $M_{Z_R}$ & 4.3 & 4.2 & 4.3 \\
        \hline
        \hline
        $W_R \to j j$, anti-diag. & \cite{ATLAS:2019fgd} & \multirow{2}{*}{$M_{W_R}$} & 2.1 & 2.0 & 2.1 \\
        \cline{4-6}
        $W_R \to \ell_i \bar{\nu}_R$, anti-diag. & \cite{ATLAS:2019lsy} & & 4.3 & 4.2 & 4.3 \\
        \hline
        \hline
        $W_R \to j j$, diag. & \cite{ATLAS:2019fgd} & \multirow{2}{*}{$M_{W_R}$} & 2.9 & 2.7 & 3.0 \\
        \cline{4-6}
        $W_R \to \ell_i \bar{\nu}_R$, diag. & \cite{ATLAS:2019lsy} & & 5.1 & 5.0 & 5.1 \\
        \hline
    \end{tabular}
    \caption{Summary of the direct bounds on the new gauge boson masses $ M_{Z_R} $ or $ M_{W_R} $ (given in TeV), according to the case indicated in the first column. 
    The more conservative bounds on $M_{W_R}$ are obtained with an anti-diagonal RH quark-mixing matrix. For comparison, we show also the stronger limits emerging from a diagonal (i.e., CKM-like) texture.
    The case $W_R \to \ell_i \bar{\nu}_R$ concerns light RH neutrinos. Bounds combine information from electronic and muonic modes.}
    \label{tab:summary_bounds}
\end{table}


We now perform a detailed discussion of indirect bounds derived from
the use of Eqs.~\eqref{ecu.2.13}, \eqref{ecu.2.18} and \eqref{ecu.2.23}
as a function of the couplings of the theory. 
Namely, in the two cases having doublet fields:

\begin{equation}\label{eq:relation_masses_doublets}
    M_{W_R} = \cos \gamma \, M_{Z_R} \,,
\end{equation}
while in the model with triplets

\begin{equation}\label{eq:relation_masses_triplets}
    \sqrt{2} \, M_{W_R} = \cos \gamma \, M_{Z_R} \,.
\end{equation}

\noindent
As seen from these expressions, the masses of the \(W_R\) and \(Z_R\) gauge bosons are related by \(\cos\gamma\), where $\gamma$ is defined in Eq.~\eqref{ecu.2.3}.
Consequently, by imposing constraints on one of these masses, we can establish limits on the mass of the other gauge boson, taking into account the permissible range of \(\gamma\) as determined by perturbativity considerations.
In scenarios (D), (T) and (Eff), we have that $ M_{W_R} \leq M_{Z_R} $.\footnote{As such, Eqs.~\eqref{eq:relation_masses_doublets} or \eqref{eq:relation_masses_triplets} would not hold, for instance, under the breaking pattern $ SU(2)_R \times U(1)_X \to U(1)_{3R} \times U(1)_X \to U(1)_Y $ \cite{Mohapatra:1998rq}, not considered in this work. In the latter case, one could still exploit that $ M_{W_R} > M_{Z_R} $, see also Ref.~\cite{Mohapatra:1986uf}.}

The use of these relations is illustrated with dotted orange curves in
Fig.~\ref{fig:direct_indirect_mass_bounds}.
Note that the expression relating the masses $ M_{Z_R} $ and $ M_{W_R} $ depends on the specific scenario analysed, as shown in Eqs.~\eqref{eq:relation_masses_doublets} and \eqref{eq:relation_masses_triplets}.
Consequently, the indirect constraints vary according to whether the doublet or triplet realizations of the scalar sector are considered.
Depending of the value of $\gamma$ inside the perturbative region, the ratio of direct vs. indirect bounds on the gauge boson masses from searches in colliders can vary by as much as a factor of 2. In particular, this is the case for the LR symmetric scenario (\(\gamma\approx33^{\circ}\)) for the mass of the \(W_R\) boson in the effective model.
Fig.~\ref{fig:direct_indirect_mass_bounds} also demonstrates that for lower (higher) values of $\gamma$, we can achieve up to about a 50\% improvement in the bounds on the $W_R$ (respectively, $Z_R$) gauge boson mass.
This highlights the complementarity of the two strategies in the context of the Doublet, Triplet and Effective LR models.

We have used the data corresponding to the narrow-width approximation for all the values of \(\gamma\). As stated in Sec.~\ref{sec:direct_bounds_Zprime}, considering different values of the \(Z_R\) width
in the range that corresponds to our case (\(\Gamma_{Z_R}/M_{Z_R}\) between 2\% and 10\%) has little impact on the bounds on its mass. On the other hand,
\(\Gamma_{W_R}/M_{W_R}\lesssim5\%\) when \(\gamma>30^{\circ}\);
small discrepancies with respect to the narrow width approximation could then only be found for smaller values of the mixing angle, i.e., \(\gamma<30^{\circ}\). However, the bounds for small values of $\gamma$ are dominated by the indirect bounds derived from $Z_R$ production.





\begin{figure}
    \centering
    \includegraphics[scale=0.29]{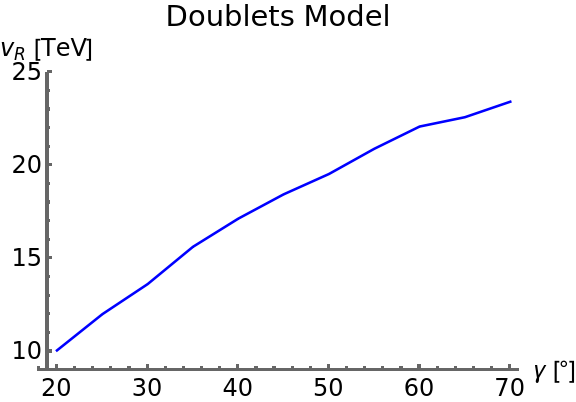} \hspace{5mm}
    \includegraphics[scale=0.29]{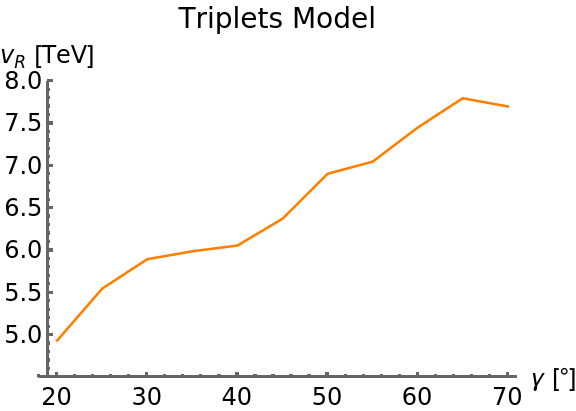} \hspace{5mm}
    \includegraphics[scale=0.29]{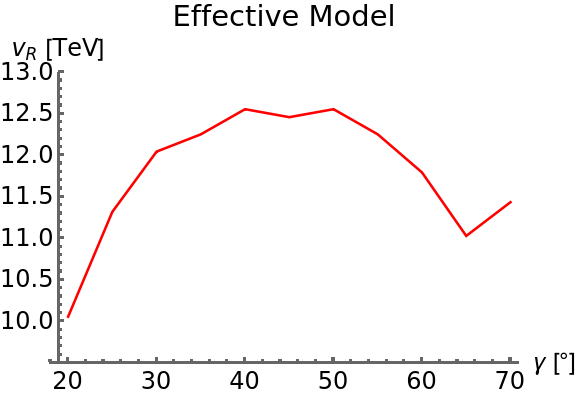}
    \caption{Bounds on the LR scale 
    in the doublet (left panel, in blue), triplet (middle panel, in orange) and effective (right panel, in red) models, as functions of the mixing angle \(\gamma\).
    }
    \label{fig:vR:bounds}
\end{figure}

As seen from the dashed red curves in Fig.~\ref{fig:direct_indirect_mass_bounds}, combining direct and indirect bounds, masses are constrained to lie above: (D) $M_{W_R} > 4.3$~TeV and $M_{Z_R} > 5.4$~TeV in the case of the ``bidoublet + two doublets'' model; (T) $M_{W_R} > 2.0$~TeV and $M_{Z_R} > 4.2$~TeV in the case of the ``bidoublet + two triplets'' model; (Eff) $M_{W_R} > 2.1$~TeV and $M_{Z_R} > 4.3$~TeV in the case of the effective ``two doublets'' model. In the first (D) case, neutrinos are necessarily of Dirac kind, and thus of negligible masses, ultimately implying the larger values for the bounds.
Given the values of the lower bounds on the gauge-boson masses as a function of $\gamma$, and the perturbativity requirement on the gauge couplings, we extract 
the lower bounds on the LR energy scale shown in Fig.~\ref{fig:vR:bounds}. The less constraining limits are obtained for the lowest allowed value of the mixing angle 
\(\gamma=20^{\circ}\), which results in the absolute bounds:
%
\begin{equation}
    |v_R|_D \gtrsim 10\,\mathrm{TeV},\hspace{0.6cm} |v_R|_T \gtrsim 4.9\,\mathrm{TeV},\hspace{0.6cm} |v_R|_{\rm Eff} \gtrsim 10\,\mathrm{TeV}.
\end{equation}

To conclude this section, let us comment on the use of a light scalar spectrum to maximize the $Z_R$ and $W_R$ total widths.
Other observables, such as those stemming from flavour physics, tend to push the masses of the scalar sector to large values. The upper limits of the partial widths would then not be achieved, contrarily to what has been considered in this paper, and this would introduce model-dependence in the calculation of the partial widths. The branching ratios for fermionic decays of the LRM gauge bosons would increase, and the lower bounds on their masses would thus also increase, in a way depending on the specific construction of each LRM scalar realization. This would, however, only slightly increase the bounds provided.


\section{Conclusions}\label{sec:conclusions}


LRMs provide the possibility of restoring discrete symmetries at high energies, giving an explanation for their violation at low energies.
Different realizations of their scalar
sectors are possible, which involve a large number of new parameters in the scalar potential.
Moreover, apart from a new gauge coupling, the model introduces new fermion mixing matrices.


We illustrate the impact of the scalar sector by considering three scenarios: (D) a model with two doublets and a bi-doublet; (T) a model with two triplets and a bi-doublet; and (Eff) an effective model with two doublets and no bi-doublet.
Scenario (Eff) allows a straightforward analysis of LRMs by introducing the minimal scalar content necessary for the SSB of the new gauged symmetries.
Based on the use of the equivalence theorem, relating longitudinal vector boson degrees of freedom to would-be Goldstone modes in the large boosted regime, the total widths from two-body decays of the new heavy gauge bosons for a fixed $SU(2)_R$ gauge coupling are maximized;
this procedure has the advantage of achieving bounds that do not depend any further on the details of the specific scalar realizations.
We also consider more general
textures of $V_R^{\mathrm{CKM}}$ with non-manifest LR symmetry,
which affect the production mechanism of the heavy charged gauge boson, due to the different PDFs of the constituent partons of the colliding protons at LHC.
The anti-diagonal texture of the RH mixing matrix leads to the less constraining bound on the mass of the $W_R$;
for comparison, we also provide bounds for the diagonal structure.
By also considering the value of the new gauge coupling, adjusted to minimize the production cross section while satisfying perturbativity constraints, we get the model-independent bounds summarized in Tab.~\ref{tab:summary_bounds} for $Z_R$ and $W_R$,
which consist of the less constraining limits on their masses.
Bounds for distinct values of the LRM gauge couplings are provided in Fig.~\ref{fig:direct_indirect_mass_bounds}.
In some cases better bounds can be achieved by exploiting the relation between the $Z_R$ and $W_R$ gauge boson masses in a particular model, showing the complementarity of neutral and charged heavy gauge-boson direct searches.

In the future, we plan to address electroweak precision bounds for different LRM realizations, which are also independent of the mixing matrices for quarks and leptons, and flavour observables, in order to further constrain distinct versions of LRMs.


\vspace{3mm}
\noindent
\textbf{Acknowledgements.}
We thank Prasanna Kumar Dhani, Andreas Hinzmann, Greg Landsberg, Jeongeun Lee, Emanuela Musumeci, and Jos\'{e} Zurita for engaging in discussions and providing their comments. We also thank Chitrangada Devi for carefully reading the manuscript.

\vspace{3mm}
\noindent
This work has been supported by MCIN/AEI/10.13039/501100011033, grant PID2020-114473GB-I00; by Generalitat Valenciana, grant PROMETEO/2021 /071 and by Ministerio de Universidades (Gobierno de Espa\~na), grant FPU20/04279.
This project has received funding from the European Union’s Horizon 2020 research and innovation programme under the Marie Sklodowska-Curie grant agreement No 101031558.
LVS is grateful for the hospitality of the CERN-TH group where part of this research was executed.

\newpage
\appendix

\section{Scalar potentials}\label{sec:Scalar_Potentials}

Here we display the explicit form of the scalar potentials in the most general case where no additional discrete symmetry is imposed to the theory. The only difference with respect to the LR symmetric scenario is that some of the parameters of the potential are no longer real. Thus, they are sources of \(\mathcal{CP}\) violation. For the models with doublets and triplets we have, respectively:

\begin{equation}
\label{ecu.B.1}
    \begin{split}
    V_D&=-\mu_1^2\,\mathrm{tr}\left\{\Phi\Phi^{\dagger}\right\}+\left(-\mu_2^2\,\mathrm{tr}\left\{\tilde{\Phi}\Phi^{\dagger}\right\}+\mathrm{h.c.}\right)-\mu^2_L\,\chi_L^{\dagger}\chi_L-\mu^2_R\,\chi_R^{\dagger}\chi_R\\&
    +\lambda_1\,\mathrm{tr}\left\{\Phi\Phi^{\dagger}\right\}^2+\lambda_2\,\mathrm{tr}\left\{\tilde{\Phi}\Phi^{\dagger}\right\}\mathrm{tr}\left\{\Phi\tilde{\Phi}^{\dagger}\right\}
    \\&+\left(\lambda_3\,\mathrm{tr}\left\{\tilde{\Phi}\Phi^{\dagger}\right\}^2+\lambda_4\,\mathrm{tr}\left\{\Phi\Phi^{\dagger}\right\}\mathrm{tr}\left\{\tilde{\Phi}\Phi^{\dagger}\right\}+\mathrm{h.c.}\right)\\&+
    \lambda_L\left(\chi_L^{\dagger}\chi_L\right)^2+\lambda_R\left(\chi_R^{\dagger}\chi_R\right)^2+\lambda_{LR}\left(\chi_L^{\dagger}\chi_L\right)\left(\chi_R^{\dagger}\chi_R\right) \\& +\left(\mu_1'\chi_L^{\dagger}\Phi\chi_R+\mu_2'\chi_L^{\dagger}\tilde{\Phi}\chi_R+\mathrm{h.c.}\right)\\& +\alpha_{1,L}\,\mathrm{tr}\left\{\Phi\Phi^{\dagger}\right\}\chi_L^{\dagger} \chi_L+\left(\alpha_{2,L}\,\mathrm{tr}\left\{\tilde{\Phi}\Phi^{\dagger}\right\}\chi_L^{\dagger}\chi_L+\mathrm{h.c.}\right)
    \\&+\alpha_{3,L}\,\chi_L^{\dagger}\Phi\Phi^{\dagger}\chi_L+\alpha_{4,L}\,\chi_L^{\dagger}\tilde{\Phi}\tilde{\Phi}^{\dagger}\chi_L
    \\&+\alpha_{1,R}\,\mathrm{tr}\left\{\Phi\Phi^{\dagger}\right\}\chi_R^{\dagger}\chi_R+\left(\alpha_{2,R}\,\mathrm{tr}\left\{\tilde{\Phi}\Phi^{\dagger}\right\}\chi_R^{\dagger}\chi_R+\mathrm{h.c.}\right)
    \\&+\alpha_{3,R}\,\chi_R^{\dagger}\Phi^{\dagger}\Phi\chi_R+\alpha_{4,R}\,\chi_R^{\dagger}\tilde{\Phi}^{\dagger}\tilde{\Phi}\chi_R
    ,
    \end{split}
\end{equation}

\begin{equation}
\label{ecu.B.2}
    \begin{split}
    V_T&=-\mu_1^2\,\mathrm{tr}\left\{\Phi\Phi^{\dagger}\right\}-\left(\mu_2^2\,\mathrm{tr}\left\{\tilde{\Phi}\Phi^{\dagger}\right\}+\mathrm{h.c.}\right)-\mu^2_L\,\mathrm{tr}\left\{\Delta_L\Delta_L^{\dagger}\right\}-\mu^2_R\,\mathrm{tr}\left\{\Delta_R\Delta_R^{\dagger}\right\}\\&
    +\lambda_1\,\mathrm{tr}\left\{\Phi\Phi^{\dagger}\right\}^2+\lambda_2\,\mathrm{tr}\left\{\tilde{\Phi}\Phi^{\dagger}\right\}\mathrm{tr}\left\{\Phi\tilde{\Phi}^{\dagger}\right\}
    \\&+\left(\lambda_3\,\mathrm{tr}\left\{\tilde{\Phi}\Phi^{\dagger}\right\}^2+\lambda_4\,\mathrm{tr}\left\{\Phi\Phi^{\dagger}\right\}\mathrm{tr}\left\{\tilde{\Phi}\Phi^{\dagger}\right\}+\mathrm{h.c.}\right)\\&
    +\rho_{1,L}\,\mathrm{tr}\left\{\Delta_L\Delta_L^{\dagger}\right\}^2+\rho_{2,L}\,\mathrm{tr}\left\{\Delta_L\Delta_L\right\}\mathrm{tr}\left\{\Delta_L^{\dagger}\Delta_L^{\dagger}\right\}\\&+\rho_{1,R}\,\mathrm{tr}\left\{\Delta_R\Delta_R^{\dagger}\right\}^2+\rho_{2,R}\,\mathrm{tr}\left\{\Delta_R\Delta_R\right\}\mathrm{tr}\left\{\Delta_R^{\dagger}\Delta_R^{\dagger}\right\}\\& +\rho_3\,\mathrm{tr}\left\{\Delta_L\Delta_L^{\dagger}\right\}\left\{\Delta_R\Delta_R^{\dagger}\right\}+\left(\rho_4\,\mathrm{tr}\left\{\Delta_L\Delta_L\right\}\left\{\Delta_R^{\dagger}\Delta_R^{\dagger}\right\}+\mathrm{h.c.}\right) \\&+\alpha_{1,L}\,\mathrm{tr}\left\{\Phi\Phi^{\dagger}\right\}\mathrm{tr}\left\{\Delta_L\Delta_L^{\dagger}\right\}+\left(\alpha_{2,L}\,\mathrm{tr}\left\{\tilde{\Phi}\Phi^{\dagger}\right\}\mathrm{tr}\left\{\Delta_L\Delta_L^{\dagger}\right\}+\mathrm{h.c.}\right)
    \\&+\alpha_{1,R}\,\mathrm{tr}\left\{\Phi\Phi^{\dagger}\right\}\mathrm{tr}\left\{\Delta_R\Delta_R^{\dagger}\right\}+\left(\alpha_{2,R}\,\mathrm{tr}\left\{\tilde{\Phi}\Phi^{\dagger}\right\}\mathrm{tr}\left\{\Delta_R\Delta_R^{\dagger}\right\}+\mathrm{h.c.}\right)
    \\&+\alpha_{3,L}\,\mathrm{tr}\left\{\Phi\Phi^{\dagger}\Delta_L\Delta_L^{\dagger}\right\}
    +\alpha_{3,R}\,\mathrm{tr}\left\{\Phi^{\dagger}\Phi\Delta_R\Delta_R^{\dagger}\right\}\\& + \left(\beta_1\mathrm{tr}\left\{\Phi\Delta_R\Phi^{\dagger}\Delta_L^{\dagger}\right\}+\beta_2\mathrm{tr}\left\{\tilde{\Phi}\Delta_R\Phi^{\dagger}\Delta_L^{\dagger}\right\}+\beta_3\mathrm{tr}\left\{\Phi\Delta_R\tilde{\Phi}^{\dagger}\Delta_L^{\dagger}\right\}+\mathrm{h.c.}\right),
    \end{split}
\end{equation}

\noindent where all of the parameters are real if the addition of the hermitian conjugate for the associated term is not indicated.



The fact that $ \tilde{\Phi}\Phi^{\dagger} $ is diagonal was used to simplify the potentials.
In the potential $V_T$ single-trace terms with four triplet fields have been eliminated with the $SU(2)$ algebraic identity ($x=x^i \sigma^i$; $x=a,b,c,d$)
\begin{equation}
2\,\mathrm{tr}\left\{ abcd\right\}\, =\,  
\mathrm{tr}\left\{ ab\right\}\mathrm{tr}\left\{ cd\right\}
-\mathrm{tr}\left\{ ac\right\}\mathrm{tr}\left\{ bd\right\}
+\mathrm{tr}\left\{ ad\right\}\mathrm{tr}\left\{ bc\right\} .
\end{equation}

\section{Summary of couplings}\label{sec:summary_couplings}

Here we display all of the terms of the Lagrangian that we have used to calculate the widths and cross sections of the different $1 \to 2$ processes. We have decomposed the fields as \(\chi_{L,R}^0\coloneqq\left(v_{L,R}+\chi_{L,R}^{0r}+i\,\chi_{L,R}^{0i}\right)/\sqrt{2}\), \(\delta_{L,R}^0\coloneqq\left(v_{L,R}+\delta_{L,R}^{0r}+i\,\delta_{L,R}^{0i}\right)/\sqrt{2}\) and \(\phi_{1,2}^0\coloneqq\left(\kappa_{1,2}+\phi_{1,2}^{0r}\pm\,i\,\phi_{1,2}^{0\,i}\right)/\sqrt{2}\).
All of the couplings are given at leading order in powers of \(v_{EW}/v_R\).

\begin{itemize}
    \item \(Z_R\) with two fermions: 
    \begin{equation}
    \label{ecu.A.1}
        \begin{split}
        &\frac{1}{12}\frac{e}{\cos\theta_W}\tan\gamma\,Z_R^{\mu}\,\bar{u}\gamma_{\mu}\left\{\left(2-3\cot^2\gamma\right)-\left(3\cot^2\gamma\right)\gamma_5\right\}u
        \\& +\frac{1}{12}\frac{e}{\cos\theta_W}\tan\gamma\,Z_R^{\mu}\,\bar{d}\gamma_{\mu}\left\{\left(2+3\cot^2\gamma\right)+\left(3\cot^2\gamma\right)\gamma_5\right\}d
        \\& -\frac{1}{4}\frac{e}{\cos\theta_W}\tan\gamma\,Z_R^{\mu}\,\bar{e}\gamma_{\mu}\left\{\left(2-\cot^2\gamma\right)-\left(\cot^2\gamma\right)\gamma_5\right\}e
        \\& -\frac{1}{2}\frac{e}{\cos\theta_W}\tan\gamma\, Z_R^{\mu}\,\bar{\nu}_L\gamma_{\mu}\nu_L-\frac{1}{2}\frac{e}{\cos\theta_W}\tan\gamma\left(1+\cot^2\gamma\right)\,Z_R^{\mu}\,\bar{\nu}_R\gamma_{\mu}\nu_R.
        \end{split}
    \end{equation}
    
    \item \(Z_R\) with two neutral scalars in the models with doublets: 
    \begin{equation}
    \label{ecu.A.2}
        \begin{split}
        Z_R^{\mu}&\left\{\frac{1}{2}\,g_R\cos\gamma\left[\phi_1^{0i}\partial_{\mu}\phi_1^{0r}-\phi_1^{0r}\partial_{\mu}\phi_1^{0i}+\phi_2^{0i}\partial_{\mu}\phi_2^{0r}-\phi_2^{0r}\partial_{\mu}\phi_2^{0i}\right]\right.\\&+\frac{1}{2}\,g_{X}\sin\gamma\left[\chi_L^{0i}\partial_{\mu}\chi_L^{0r}-\chi_L^{0r}\partial_{\mu}\chi_L^{0i}\right]\\&\left.+\frac{1}{2}\,\left(g_R\cos\gamma+g_{X}\sin\gamma\right)\left[\chi_R^{0i}\partial_{\mu}\chi_R^{0r}-\chi_R^{0r}\partial_{\mu}\chi_R^{0i}\right]\right\}.
        \end{split}
    \end{equation}
    
    \item \(Z_R\) with two neutral scalars in the models with triplets: 
    \begin{equation}
    \label{ecu.A.3}
        \begin{split}
        Z_R^{\mu}&\left\{\frac{1}{2}\,g_R\cos\gamma\left[\phi_1^{0i}\partial_{\mu}\phi_1^{0r}-\phi_1^{0r}\partial_{\mu}\phi_1^{0i}+\phi_2^{0i}\partial_{\mu}\phi_2^{0r}-\phi_2^{0r}\partial_{\mu}\phi_2^{0i}\right]\right.\\&+\,g_{X}\sin\gamma\left[\delta_L^{0i}\partial_{\mu}\delta_L^{0r}-\delta_L^{0r}\partial_{\mu}\delta_L^{0i}\right]\\&\left.+\left(g_R\cos\gamma+g_{X}\sin\gamma\right)\left[\delta_R^{0i}\partial_{\mu}\delta_R^{0r}-\delta_R^{0r}\partial_{\mu}\delta_R^{0i}\right]\right\}.
        \end{split}
    \end{equation}

    \item \(Z_R\) with two singly charged scalars in the models with doublets: 
    \begin{equation}
    \label{ecu.A.4}
        \begin{split}
        i\,Z_R^{\mu}&\left\{\frac{1}{2}\,g_R\cos\gamma\left[\phi_1^{+}\partial_{\mu}\phi_1^{-}-\phi_1^{-}\partial_{\mu}\phi_1^{+}+\phi_2^{+}\partial_{\mu}\phi_2^{-}-\phi_2^{-}\partial_{\mu}\phi_2^{+}\right]\right.\\&+\frac{1}{2}\,g_{X}\sin\gamma\left[\chi_L^{-}\partial_{\mu}\chi_L^{+}-\chi_L^{+}\partial_{\mu}\chi_L^{-}\right]\\&\left.+\frac{1}{2}\left(g_R\cos\gamma-g_{X}\sin\gamma\right)\left[\chi_R^{+}\partial_{\mu}\chi_R^{-}-\chi_R^{-}\partial_{\mu}\chi_R^{+}\right]\right\}.
        \end{split}
    \end{equation}

    \item \(Z_R\) with two singly charged scalars in the models with triplets: 
    \begin{equation}
    \label{ecu.A.5}
        \begin{split}
        i\,Z_R^{\mu}&\left\{\frac{1}{2}\,g_R\cos\gamma\left[\phi_1^{+}\partial_{\mu}\phi_1^{-}-\phi_1^{-}\partial_{\mu}\phi_1^{+}+\phi_2^{+}\partial_{\mu}\phi_2^{-}-\phi_2^{-}\partial_{\mu}\phi_2^{+}\right]\right.\\&+g_{X}\sin\gamma\left[\delta_L^{-}\partial_{\mu}\delta_L^{+}-\delta_L^{+}\partial_{\mu}\delta_L^{-}+\delta_R^{-}\partial_{\mu}\delta_R^{+}-\delta_R^{+}\partial_{\mu}\delta_R^{-}\right]\bigg\}.
        \end{split}
    \end{equation}

    \item \(Z_R\) with two doubly charged scalars: 
    \begin{equation}
    \label{ecu.A.6}
    \begin{split}
    i\,Z^{\mu}_R&\left\{\left(g_R\cos\gamma-g_X\sin\gamma\right)\left[\delta_R^{++}\partial_{\mu}\delta_R^{--}-\delta_R^{--}\partial_{\mu}\delta_R^{++}\right]\right.\\&\left.-g_X\sin\gamma\left[\delta_L^{++}\partial_{\mu}\delta_L^{--}-\delta_L^{--}\partial_{\mu}\delta_L^{++}\right]\right\}.
    \end{split}
    \end{equation}

    \item \(Z_R\) with \(W_R\) and a singly charged scalar in the models with doublets: 
    \begin{equation}
    \label{ecu.A.7}
        -\frac{e}{\cos\theta_W}\sin\gamma M_{Z_R}\chi_R^{-}W_{R,\mu}^+Z_R^{\mu}+\mathrm{h.c.}
    \end{equation}
     
    \item \(Z_R\) with \(W_R\) and a singly charged scalar in the models with triplets: 
    \begin{equation}
        \label{ecu.A.8}
        -\frac{1}{\sqrt{2}}\frac{e}{\cos\theta_W}\sin\gamma\left(2+\cot^2\gamma\right)M_{Z_R}\delta_R^{-}W_{R,\mu}^+Z_R^{\mu}+\mathrm{h.c.}
    \end{equation}

    \item \(Z_R\) with two \(W_L\) bosons: 
    \begin{equation}
        \label{ecu.A.9}
        \begin{split}
        i e \cot \theta_W \sin \alpha &\left\{(\partial^{\mu}W_L^{-\nu})(W_{L,\mu}^+Z_{R,\nu}-W_{L,\nu}^+Z_{R,\mu})+(\partial^{\mu}W_L^{+\nu})(W_{L,\nu}^-Z_{R,\mu}-W_{L,\mu}^-Z_{R,\nu})\right.\\&\left.+(\partial^{\mu}Z_R^{\nu})(W_{L,\mu}^-W_{L,\nu}^+-W_{L,\nu}^-W_{L,\mu}^+)\right\},
    \end{split}
    \end{equation}
    where $ \sin \alpha = \frac{\tan\gamma \tan \theta_W}{\cos \theta_W}\frac{M_{W_L}^2}{M_{Z_R}^2} $ results from the mixing of the massive neutral gauge bosons 
    in the \(\chi_L+\chi_R\) Effective LR Model.

    \item \(Z_R\) with two \(W_R\) bosons: 
    \begin{equation}
        \label{ecu.A.10}
        \begin{split}
        i\frac{e}{\cos\theta_W}\cot\gamma&\left\{(\partial^{\mu}W_R^{-\nu})(W_{R,\mu}^+Z_{R,\nu}-W_{R,\nu}^+Z_{R,\mu})+(\partial^{\mu}W_R^{+\nu})(W_{R,\nu}^-Z_{R,\mu}-W_{R,\mu}^-Z_{R,\nu})\right.\\&\left.+(\partial^{\mu}Z_R^{\nu})(W_{R,\mu}^-W_{R,\nu}^+-W_{R,\nu}^-W_{R,\mu}^+)\right\}.
    \end{split}
    \end{equation}

    \item \(W_R\) with two fermions (Dirac neutrinos): 
    \begin{equation}
    \label{ecu.A.11}
    -\frac{g_R}{\sqrt{2}}W_R^{+\,\mu}\left\{\bar{u}\gamma_{\mu}V_R^{\mathrm{CKM}}\mathcal{P}_Rd+\bar{\nu}\gamma_{\mu}V_R^{\mathrm{PMNS}}\mathcal{P}_Re\right\}+\mathrm{h.c.}
    \end{equation}

    \item \(W_R\) with two fermions (Majorana neutrinos): 
    \begin{equation}
    \label{ecu.A.12}
    -\frac{g_R}{\sqrt{2}}W_R^{+\,\mu}\left\{\bar{u}\gamma_{\mu}V_R^{\mathrm{CKM}}\mathcal{P}_Rd+\bar{\nu}_h\gamma_{\mu}V_h^{\textrm{PMNS}}\mathcal{P}_R e
    \right\}+\mathrm{h.c.}
    \end{equation}

    \item \(W_R\) with a singly charged and a neutral scalar in the models with doublets: 
    \begin{equation}
    \label{ecu.A.13}
    \begin{split}
    &\frac{g_R}{2}W_R^{+\mu}\left\{\phi_1^{0i}\partial_{\mu}\phi_1^--\phi_1^-\partial_{\mu}\phi_1^{0i}+\phi_2^-\partial_{\mu}\phi_2^{0i}-\phi_2^{0i}\partial_{\mu}\phi_2^-+\chi_R^-\partial_{\mu}\chi_R^{0i}-\chi_R^{0i}\partial_{\mu}\chi_R^-\right.\\ &\left.+i\left[\phi_1^-\partial_{\mu}\phi_1^{0r}-\phi_1^{0r}\partial_{\mu}\phi_1^-+\phi_2^{0r}\partial_{\mu}\phi_2^--\phi_2^-\partial_{\mu}\phi_2^{0r}+\chi_R^{0r}\partial_{\mu}\chi_R^--\chi_R^-\partial_{\mu}\chi_R^{0r}\right]\right\}+\mathrm{h.c.}
    \end{split}
    \end{equation}

    \item \(W_R\) with a singly charged and a neutral scalar in the models with triplets: 
    \begin{equation}
    \label{ecu.A.14}
    \begin{split}
    &\frac{g_R}{2}W_R^{+\mu}\left\{\phi_1^{0i}\partial_{\mu}\phi_1^--\phi_1^-\partial_{\mu}\phi_1^{0i}+\phi_2^-\partial_{\mu}\phi_2^{0i}-\phi_2^{0i}\partial_{\mu}\phi_2^-+\sqrt{2}\left(\delta_R^-\partial_{\mu}\delta_R^{0i}-\delta_R^{0i}\partial_{\mu}\delta_R^-\right)\right.\\ &\left.+i\left[\phi_1^-\partial_{\mu}\phi_1^{0r}-\phi_1^{0r}\partial_{\mu}\phi_1^-+\phi_2^{0r}\partial_{\mu}\phi_2^--\phi_2^-\partial_{\mu}\phi_2^{0r}+\sqrt{2}\left(\delta_R^{0r}\partial_{\mu}\delta_R^--\delta_R^-\partial_{\mu}\delta_R^{0r}\right)\right]\right\}+\mathrm{h.c.}
    \end{split}
    \end{equation}

     \item \(W_R\) with a doubly and a singly charged scalars: 
    \begin{equation}
        \label{ecu.A.15}
        ig_RW_R^{+\mu}\left\{\delta_R^{--}\partial_{\mu}\delta_R^+-\delta_R^+\partial_{\mu}\delta_R^{--}\right\}+\mathrm{h.c.}
    \end{equation}

    \item \(W_R\) with \(W_L\) and a doubly charged scalar: 
    \begin{equation}
        \label{ecu.A.16}
        \sqrt{2}g_R^2v_R e^{i\lambda}\sin\xi\cos\xi W_R^{-\mu}W_{L\mu}^-\delta_R^{++}+\mathrm{h.c.}
    \end{equation}

    \item \(W_R\) with a singly charged scalar and a photon in the model with a bidoublet and two doublets (exact result up to \(W_L-W_R\) mixing coefficients): 
    \begin{equation}
        \label{ecu.A.17}
        \frac{1}{2}eg_RW_R^{+\mu}A_{\mu}\left\{v_R\chi_R^{-}+\kappa_2^{\ast}\phi_2^--\kappa_1\phi_1^-\right\}+\mathrm{h.c.}
    \end{equation}
    \item \(W_R\) with a singly charged scalar and a photon in the model with a bidoublet and two triplets (exact result up to \(W_L-W_R\) mixing coefficients): 
    \begin{equation}
        \label{ecu.A.18}
         \frac{1}{2}eg_RW_R^{+\mu}A_{\mu}\left\{\sqrt{2}v_R\delta_R^{-}+\kappa_2^{\ast}\phi_2^--\kappa_1\phi_1^-\right\}+\mathrm{h.c.}
    \end{equation}
    \item \(W_R\) with scalars in the model with a bidoublet and two doublets (exact result up to \(W_L-W_R\) mixing coefficients): 
    \begin{equation}
        \label{ecu.A.19}
        \frac{i}{2}g_RW_R^{+\mu}\partial_{\mu}\left\{v_R\chi_R^{-}+\kappa_2^{\ast}\phi_2^--\kappa_1\phi_1^-\right\}+\mathrm{h.c.}
    \end{equation}
    \item \(W_R\) with scalars in the model with a bidoublet and two triplets (exact result up to \(W_L-W_R\) mixing coefficients): 
    \begin{equation}
        \label{ecu.A.20}
        \frac{i}{2}g_RW_R^{+\mu}\partial_{\mu}\left\{\sqrt{2}v_R\delta_R^{-}+\kappa_2^{\ast}\phi_2^--\kappa_1\phi_1^-\right\}+\mathrm{h.c.}
    \end{equation}
\end{itemize}

Eqs.~\eqref{ecu.A.19} and \eqref{ecu.A.20} are used to identify the combinations of scalar degrees of freedom that result being the Goldstone bosons, see App.~\ref{sec:Non-Fermionic_Decays}.

There is no interaction vertex between \(W_R\), \(W_L\) and \(A\) at the tree level; we do have couplings of the form \(W_1W_1A\) and \(W_2W_2A\) (while \(W_1 W_2 A\) is not possible) but when we apply the transformation in Eq.~\eqref{ecu.2.1} the terms that contain the vertex \(W_LW_RA\) cancel each other due to the unitarity of the mixing matrix.

Neither cubic nor quartic interaction vertices among only neutral gauge bosons can be present. Indeed, the gauge self-interactions are generated by terms proportional to \(\mathrm{tr}\left\{\partial^{\mu}W^{\nu}\left[W_{\mu},W_{\nu}\right]\right\}\) and \(\mathrm{tr}\left\{\left[W^{\mu},W^{\nu}\right]\left[W_{\mu},W_{\nu}\right]\right\}\), but the commutator \(\left[W_{\mu},W_{\nu}\right]\) does not contain products of the form \(W_{\mu}^3W_{\nu}^3\) since \(\left[\sigma^3,\sigma^3\right]=0\).



\section{Non-fermionic decays of the \(Z_R\) and \(W_R\) gauge bosons}\label{sec:Non-Fermionic_Decays}

Here we show the details of the calculation of the bounds on the partial widths of the non-fermionic decays of the \(W_R\) and \(Z_R\) gauge bosons. We start with the latter. First of all, let us calculate the partial width of the process \(Z_R\) decaying to two particular neutral scalars. The Lagrangian of the interaction can be written as

\begin{equation}
    \label{ecu.C.1}
    \mathcal{L}=Z_R^{\mu}\sum_{a,\,b}C_{ab}\left\{\psi_a\partial_{\mu}\psi_b-\psi_b\partial_{\mu}\psi_a\right\},
\end{equation}

\noindent where the \(C_{ab}\) are the coupling constants and \(\left\{\psi_a\right\}_{a=1}^{8}\) is just a basis of the neutral fields. Looking at Eqs.~(\ref{ecu.A.2}) and (\ref{ecu.A.3}) we see that Eq.~(\ref{ecu.C.1}) can be used to describe both the cases of doublets and triplets with the bi-doublet. In the interaction basis the non-vanishing couplings are \(C_{12}^D=C_{12}^T=C_{34}^D=C_{34}^T=\frac{1}{2}g_R\cos\gamma\), \(C_{56}^D=\frac{1}{2}\,C_{56}^T=\frac{1}{2}g_X\sin\gamma\) and \(C_{78}^D=\frac{1}{2}\,C_{78}^T=\frac{1}{2}\,\left(g_R\cos\gamma+g_X\sin\gamma\right)\). Now, we have to make a rotation into the basis of the physical fields \(\left\{\psi'_a\right\}_{a=1}^{8}\). We use the transformation  \(\psi_a=\sum_i\mathbb{R}_{ai}\psi_i'\). Then, defining the matrix \(C'\coloneqq\mathbb{R}^{\mathrm{T}}\,C\,\mathbb{R}\) we can write

\begin{equation}
    \label{ecu.C.2}
    \mathcal{L}=Z_R^{\mu}\sum_{i,\,j}C_{ij}'\left\{\psi'_i\partial_{\mu}\psi'_j-\psi'_j\partial_{\mu}\psi'_i\right\}.
\end{equation}

\noindent Therefore, the partial width of the process \(Z_R\rightarrow\psi_i'\psi_j'\) is just

\begin{equation}
    \label{ecu.C.3}
    \Gamma\left(Z_R\rightarrow\psi_i'\psi_j'\right)=\frac{M_{Z_R}}{48\pi}\left(C'_{ij}-C'_{ji}\right)^2\left[1-2\left(x_i+x_j\right)+\left(x_i-x_j\right)^2\right]^{3/2},
\end{equation}

\noindent where \(x_i\coloneqq m_i^2/M_{Z_R}^2\). It is not difficult to show that the maximum possible value of the partial width is achieved when \(x_i=x_j=0\), i.e., in the massless case. Thus, we put the following bound:

\begin{equation}
    \label{ecu.C.4}
    \Gamma\left(Z_R\rightarrow2\,\mathrm{Neutral\,Scalars}\right)<\frac{M_{Z_R}}{48\pi}\sum_{j>i}\left(C'_{ij}-C'_{ji}\right)^2=\frac{M_{Z_R}}{48\pi}\frac{1}{2}\mathrm{tr}\left\{\left(C'-C'^{\mathrm{T}}\right)\left(C'^{\mathrm{T}}-C'\right)\right\}.
\end{equation}

\noindent Using the properties of the trace and the definition of \(C'\) it is also not difficult to show that 

\begin{equation}
    \label{ecu.C.5}
    \Gamma\left(Z_R\rightarrow2\,\mathrm{Neutral\,Scalars}\right)<\frac{M_{Z_R}}{48\pi}\sum_{a,\,b}C_{ab}^2.
\end{equation}

We can do exactly the same for the decay to charged scalars. From Eqs.~(\ref{ecu.A.4}) and (\ref{ecu.A.5}) we see that we can write the interaction Lagrangian of the \(Z_R\) and 2 singly charged scalars as

\begin{equation}
    \label{ecu.C.6}
    \mathcal{L}=i\,Z_R^{\mu}\sum_aC_a\left\{\psi^+_a\partial_{\mu}\psi^-_a-\psi^-_a\partial_{\mu}\psi_a^+\right\}.
\end{equation}

\noindent
In this case, in the interaction basis we have \(C_1^D=C_1^T=C_2^D=C_2^T=\frac{1}{2}\,g_R\cos\gamma\), \(C_3^D=\frac{1}{2}\,C_3^T=\frac{1}{2}\,C_4^T=-\frac{1}{2}g_X\sin\gamma\) and \(C_4^D=\frac{1}{2}\,\left(g_R\cos\gamma-g_X\sin\gamma\right)\).
Now, we make the transformation into the physical basis \(\psi_a^{-}=\sum_i U_{ai}\psi'^{-}_i\) where \(U\) is a unitary matrix. Using the same arguments as in the previous case we can find that

\begin{equation}
    \label{ecu.C.7}
    \Gamma\left(Z_R\rightarrow2\,\mathrm{Singly\,Charged\,Scalars}\right)<\frac{M_{Z_R}}{48\pi}\sum_{i,\,j}\left|C_{ij}\right|^2,
\end{equation}

\noindent where \(C_{ij}\coloneqq\sum_a C_a U_{ai}^{\ast}U_{aj}\). Therefore, it is not difficult to prove that

\begin{equation}
    \label{ecu.C.8}
    \Gamma\left(Z_R\rightarrow2\,\mathrm{Singly\,Charged\,Scalars}\right)<\frac{M_{Z_R}}{48\pi}\sum_{a}C_a^2.
\end{equation}

If we want to calculate the bound on the partial width of the decay of the \(Z_R\) to two bosons we only need to add the contribution of \(Z_R\rightarrow\mathrm{Doubly\,Charged\,Scalars}\) in the case of the model with triplets. From Eq.~(\ref{ecu.A.6}) it is easy to see that this calculation is completely analogous to the one for singly charged scalars.

Now we will show that the partial width of the decay \(Z_R\rightarrow W_R^{\pm}W_L^{\mp}\) is negligible. The Feynman rule of the interaction of three gauge bosons is given by a function of the form \(\eta_{\mu\nu}\left(q-p\right)_{\lambda}+\eta_{\lambda\mu}\left(p-r\right)_{\nu}+\eta_{\nu\lambda}\left(r-q\right)_{\mu}\) with \(p\), \(q\) and \(r\) the incoming momenta of the particles. Then, the dependence on the mass of the partial width of the decay of one gauge boson with mass \(M\) into other two with masses \(m_1\) and \(m_2\) is given by

\begin{equation}
    \label{ecu.C.9}
    \begin{split}
    f\left(M,m_1,m_2\right)=\frac{M^5}{m_1^2m_2^2}&\left[1-2\frac{m_1^2+m_2^2}{M^2}+\left(\frac{m_1^2-m_2^2}{M^2}\right)^2\right]^{3/2}\\&\times\left\{1+10\frac{m_1^2+m_2^2}{M^2}+\frac{m_1^4+m_2^4+10m_1^2m_2^2}{M^4}\right\}.
    \end{split}
\end{equation}

\noindent In our case we can neglect the mass of the \(W_L\) boson and write

\begin{equation}
    \label{ecu.C.10}
    f\left(M_{Z_R},M_{W_R},M_{W_L}\right)\approx\frac{M^5_{Z_R}}{M_{W_L}^2M_{W_R}^2}\left(1-\frac{M^2_{W_R}}{M^2_{Z_R}}\right)^3\left[1+10\left(\frac{M_{W_R}}{M_{Z_R}}\right)^2+\left(\frac{M_{W_R}}{M_{Z_R}}\right)^4\right].
\end{equation}

\noindent
The process \(Z_R\rightarrow W_R^{\pm}W_L^{\mp}\) can only occur at the tree level if there is a mixing between the \(W_L\) and the \(W_R\) bosons. The amplitude is proportional to \(\sin\xi\), so that we have 

\begin{equation}
    \label{ecu.C.11}
    \Gamma\left(Z_R\rightarrow W_R^{\pm}W_L^{\mp}\right)\sim\frac{M^5_{Z_R}}{M_{W_L}^2M_{W_R}^2}\sin^2\xi\sim\frac{v_{EW}^2}{v_R}.
\end{equation}

\noindent
Consequently, this term can be neglected in the calculation of the width of the \(Z_R\).

We can also prove that the partial width of the decay \(Z_R\to H^{\pm}W_R^{\mp}\) is negligible. The process is absent in the \(\chi_L+\chi_R\) Effective LR Model since there are no physical charged scalars. On the other hand, as we can see in Eqs. \eqref{ecu.A.7} and \eqref{ecu.A.8}, the scalars that mediate the decay in the interaction eigenstates basis in the models with a bidoublet and two doublets or two triplets are \(\chi_R^{\pm}\) and \(\delta_R^{\pm}\), respectively. Looking at Eqs. \eqref{ecu.A.19} and \eqref{ecu.A.20} we can see that the linear combinations \(v_R\chi_R^{-}+\kappa_2^{\ast}\phi_2^--\kappa_1\phi_1^-\), \(\sqrt{2}v_R\delta_R^{-}+\kappa_2^{\ast}\phi_2^--\kappa_1\phi_1^-\) and their hermitian conjugates are proportional to a Goldstone boson (i.e., this interacting term can be cancelled by an appropriate choice of gauge-fixing Lagrangian). Consequently, the mixing of the fields \(\chi_R^{\pm}\) and \(\delta_R^{\pm}\) with physical scalars is \(\mathcal{O}\left(v_{EW}/v_R\right)\). Thus, the partial width is, at most, proportional to \(v_{EW}^2/v_R\) so it can be neglected.

Finally, we have to calculate the partial width of the decay \(Z_R\rightarrow W_R^+W_R^-\). Using the coupling shown in Eq.~(\ref{ecu.A.10}) we can see that 

\begin{equation}
    \label{ecu.C.15}
    \Gamma\left(Z_R\rightarrow W_R^+W_R^-\right)=\frac{M_{Z_R}}{192\pi}\left(\frac{e}{\cos\theta_W}\right)^2\frac{\cot^2\gamma}{x_w^2}\left(1-4x_w\right)^{3/2}\left\{1+20x_w+12x_w^2\right\}.
\end{equation}

We now shift to the non-fermionic decays of the \(W_R\) boson. As we have pointed out in the text, we can set a bound on the partial widths of these processes by calculating the upper bound on  \(\Gamma\left(W_R\rightarrow 2\,\mathrm{Scalars}\right)\). We start with the contributions with a neutral and a singly charged scalar, whose interaction Lagrangians are shown in Eqs.~(\ref{ecu.A.13}) and (\ref{ecu.A.14}). We can write them as

\begin{equation}
    \label{ecu.C.16}
    \mathcal{L}=W_R^+\sum_{ij}C_{ij}\left\{\psi_i^0\partial_{\mu}\psi_j^--\psi_j^-\partial_{\mu}\psi_i^0\right\}+\mathrm{h.c.}
\end{equation}

\noindent
We make the transformations into the mass eigenstates \(\psi_i^0=\sum_k\mathbb{R}_{ik}\psi'^{0}_k\) and \(\psi_j^-=\sum_l\mathrm{U}_{jl}\psi'^{-}_l\). Then, we have 

\begin{equation}
    \label{ecu.C.17}
    \mathcal{L}=W_R^+\sum_{ij}C'_{ij}\left\{\psi'^0_i\partial_{\mu}\psi'^-_j-\psi'^-_j\partial_{\mu}\psi'^0_i\right\}+\mathrm{h.c.},
\end{equation}

\noindent where \(C':=\mathbb{R}^{\mathrm{T}}C\mathrm{U}\). Again, the upper bound on the partial width corresponds to the case in which the particles in the final state have a negligible mass compared to the one of the \(W_R\) boson:

\begin{equation}
    \label{ecu.C.18}
    \Gamma\left(W^{\pm}_R\rightarrow\psi'^0\psi'^{\pm}\right)<\frac{M_{W_R}}{48\pi}\sum_{ij}\left|C'_{ij}\right|^2=\frac{M_{W_R}}{48\pi}\mathrm{tr}\left\{\mathbb{R}^{\mathrm{T}}C\mathrm{U}\mathrm{U}^{\dagger}C^{\dagger}\mathbb{R}\right\}=\frac{M_{W_R}}{48\pi}\sum_{ij}C_{ij}^2.
\end{equation}

The calculation for the case with a singly and a doubly charged scalars Eq.~\eqref{ecu.A.15} is completely analogous. We can also show that the contribution of \(W_R^{\pm}\rightarrow W_L^{\mp}H^{\pm\pm}\) in the models with triplets is negligible. At tree level this process is mediated by the coupling term in Eq.~\eqref{ecu.A.16}. It can be written as

\begin{equation}
    \label{ecu.C.19}
    \mathcal{L}=CW_R^{-\mu}W_{L\mu}^-\delta_R^{++}+\mathrm{h.c.}.
\end{equation}

\noindent
Since \(\sin\xi\sim v_{EW}^2/v_R^2\), then \(|C|\sim v_{EW}^2/v_R\). Now, we make a transformation into the mass eigenstates \(\delta_R^{++}=\sum_i \mathrm{U}_{2i}\psi^{++}_i\), where \(\mathrm{U}\) is a unitary transformation. Thus, we find that

\begin{equation}
    \label{ecu.C.20}
    \mathcal{L}=CW_R^{-\mu}W_{L\mu}^-\sum_i \mathrm{U}_{2i}\psi^{++}_i+\mathrm{h.c.}.
\end{equation}

\noindent Then, it can be shown that the partial width of the process \(W_R^{\pm}\rightarrow W_L^{\mp}\psi_i^{\pm\pm}\) is 

\begin{equation}
    \label{ecu.C.21}
    \Gamma\left(W_R^{\pm}\rightarrow W_L^{\mp}\psi^{\pm\pm}_i\right)\approx\frac{1}{24\pi}\frac{|C|^2}{M_{W_R}^3}\left|\mathrm{U}_{2i}\right|^2\left(M_{W_R}^2-M_i^2\right)\left\{1+\frac{\left(M_{W_R}^2-M_i^2\right)^2}{8M_{W_L}^2M_{W_R}^2}\right\},
\end{equation}

\noindent where \(M_i\) is the mass of \(\psi_i^{\pm\pm}\) and we have made the approximation \(M_{W_R}\gg M_{W_L}\). Since \(\left|\mathrm{U}_{2i}\right|^2\sim\mathcal{O}(1)\) at most, then \(\Gamma\left(W_R^{\pm}\rightarrow W_L^{\mp}\psi^{\pm\pm}_i\right)\sim v_{EW}^2/v_R\). Consequently, the partial width \(\Gamma\left(W_R^{\pm}\rightarrow W_L^{\mp}H^{\pm\pm}\right)\) is negligible.

Finally, we show that the process \(W_R^{\pm}\to H^{\pm}A\) is forbidden at tree level. Since there are no physical charged scalars in the \(\chi_L+\chi_R\) Effective LR model the result is straightforward in this case. The important interaction vertices in the models with a bidoublet and two doublets or two triplets are presented in Eqs.~\eqref{ecu.A.17} and \eqref{ecu.A.18}, respectively. Nonetheless, looking at Eqs.~\eqref{ecu.A.19} and \eqref{ecu.A.20}, it is easy to see that those combinations of scalars correspond to Goldstone bosons. This is so because of gauge invariance of the electromagnetic interaction, i.e., a vertex of the form \(W_R^{\pm\mu}A_{\mu}H^{\mp}\) with physical fields would break gauge invariance.
Consequently, there are no Feynman rules for this process with physical Higgses at tree level.

\newpage

\bibliography{mybib}{}
\bibliographystyle{unsrturl}

\end{document}